\documentclass[12pt, twoside]{article}
\usepackage[text={5.0in,8.0in},centering,top=23mm,bottom=40mm,left=30mm,right=25mm]{geometry}
\usepackage{times}
\usepackage{enumerate}

\usepackage{tabularx}

\usepackage{amsthm}
\usepackage[mathscr]{euscript}

\usepackage{pgf}
\usepackage{tikz}
\usetikzlibrary{patterns}

\usepackage[colorlinks=true,
            linkcolor=blue,
            urlcolor=blue,
            citecolor=blue]{hyperref}

\providecommand{\D}{\mathbb}

\newcommand{\ee}{\mathrm{e}}

\providecommand{\esm}[1]{\D{E}\left[ #1 \right]}

\def\dist{\mathrm{dist}}

\def\ord#1{\mathrm{o}\left(#1\right)}

\def\card{\mathrm{card}}

\def\Ran{\mathrm{Ran\,}}

\def\supp{\mathrm{supp\,}}

\def\one{\mathbf{1}}

\def\qedhere{\qed}

\definecolor{redd}{rgb}{0.95,0.2,0.2}
\definecolor{gris}{rgb}{0.9,0.9,0.9}
\definecolor{greenn}{rgb}{0.1,0.6,0.2}

\definecolor{cmgray}{rgb}{0.7,0.7,0.7}

\definecolor{cmblue}{rgb}{0.2,0.5,0.8}
\def\tmblue#1{\textcolor{cmblue}{#1}}

\def\CWk{C_{W,k}}

\DeclareMathAlphabet{\mathpzc}{OT1}{pzc}{m}{it}

\def\mes{{\mathrm{mes}}}

\def\be{\begin{equation}}
\def\ee{\end{equation}}
\def\ba{\begin{array}{l}}
\def\ea{\end{array}}

\def\bal{\begin{aligned}}
\def\eal{\end{aligned}}

\def\baln{\begin{align}}
\def\ealn{\end{align}}

\def\ble{\begin{lemma}}
\def\ele{\end{lemma}}

\def\bthm{\begin{theorem}}
\def\ethm{\end{theorem}}

\def\bco{\begin{cor}}
\def\eco{\end{cor}}

\def\bpr{\begin{prop}}
\def\epr{\end{prop}}

\def\bre{\begin{remark}}
\def\ere{\end{remark}}

\def\btm{\begin{theorem}}
\def\etm{\end{theorem}}

\def\bde{\begin{definition}}
\def\ede{\end{definition}}

\def\myset#1{\left\{ \, #1 \, \right\}}

\def\eu{{\mathrm{e}}}

\def\rPhi{{\mathrm{\Phi}}}

\def\half{\frac{1}{2}}
\def\shalf{{\textstyle{\frac{1}{2}}}}

\def\eps{\varepsilon}
\def\lam{\lambda}

\def\aL{\mathfrak{a}_{L}}
\def\bL{\mathfrak{b}_{L}}
\def\ccL{\mathfrak{c}_{L}}

\DeclareSymbolFont{newfont}{OML}{cmm}{m}{it}
\DeclareMathSymbol{\Epsilon}{3}{newfont}{15}
\DeclareMathSymbol{\Varrho}{3}{newfont}{37}
\DeclareMathSymbol{\rro}{3}{newfont}{37}

\def\pt{\partial}
\def\Const{\mathrm{Const\,}}

\def\pr#1{\mathbb{P}\left\{ #1 \right\}}
\def\bigpr#1{\mathbb{P}\big\{ #1 \big\}}

\def\esm#1{\D{E}\left[\, #1\, \right]}

\newcommand{\vertii}[1]{{\big\vert\kern-0.25ex\big\vert #1
    \big\vert\kern-0.25ex\big\vert\kern-0.25ex}}

\newcommand{\vertiii}[1]{{\big\vert\kern-0.25ex\big\vert\kern-0.25ex\big\vert #1
    \big\vert\kern-0.25ex\big\vert\kern-0.25ex\big\vert}}

\newcommand{\dnorm}[1]{{\big\| #1 \big\|^{\curlywedge}}}

\def\cell{\mathrm{C}}

\def\lam{\lambda}
\def\om{\omega}
\def\th{\theta}

\def\Lam{\Lambda}
\def\Om{\Omega}

\def\talpha{\widetilde{\alpha}}
\def\tbeta{\widetilde{\beta}}

\def\tsigma{{\widetilde{\sigma}}}
\def\tgamma{{\widetilde{\gamma}}}

\def\kcirc{{k_\circ}}

\def\vempty{\varnothing}

\def\lr#1{\langle#1\rangle}

\def\ball{\mathrm{B}}

\def\scorek{\mathbf{C}_{k}}

\def\fa{\mathfrak{a}}
\def\fb{\mathfrak{b}}

\def\fbH{\beta}

\def\fc{\mathfrak{c}}

\def\fh{\mathfrak{h}}

\def\fq{\mathfrak{q}}
\def\fs{\mathfrak{s}}
\def\fw{\mathfrak{w}}

\def\fF{\mathfrak{F}}

\def\fK{\mathfrak{K}}

\def\BA{\mathbf{A}}
\def\BB{\mathbf{B}}

\def\BF{\mathbf{F}}

\def\cY{\mathcal{Y}}

\def\csB{\mathscr{B}}

\def\csE{\mathscr{E}}

\def\cA{\mathcal{A}}
\def\cB{\mathcal{B}}
\def\cC{\mathcal{C}}

\def\cJ{\mathcal{J}}

\def\cL{\mathcal{L}}

\def\cR{\mathcal{R}}

\def\lf{\lfloor}
\def\rf{\rfloor}

\def\Lf{\left\lfloor}
\def\Rf{\right\rfloor}

\def\rc{\mathrm{c}}
\def\rd{\mathrm{d}}

\def\rC{\mathrm{C}}

\def\rGam{\mathrm{\Gamma}}
\def\rGamk{\mathrm{\Gamma}^{k}}
\def\rGamkone{\mathrm{\Gamma}^{k+1}}
\def\bGam{\boldsymbol{\rm \Gamma}}
\def\bGamk{\boldsymbol{\rm \Gamma}^{k}}
\def\bGamkone{\boldsymbol{\rm \Gamma}^{k+1}}

\def\cCk{\mathcal{C}^{k}}

\def\chik{\chi^{k}}

\def\DC{\mathbb{C}}
\def\DP{\mathbb{P}}
\def\DR{\mathbb{R}}
\def\DZ{\mathbb{Z}}
\def\DN{\mathbb{N}}

\def\tto#1{\smash{\mathop{\,\,\,\, \longrightarrow \,\,\,\, }\limits_{#1}}}

\newtheorem{theorem}{Theorem}[section]

\newtheorem{cor}{Corollary}

\newtheorem{lemma}{Lemma}

\newtheorem{definition}{Definition}

\newtheorem{remark}{Remark}

\usepackage{amssymb}
\usepackage{amsmath}
\usepackage{mathrsfs}

\usepackage{xspace}

\usepackage{epsfig}
\usepackage{enumerate}
\usepackage{graphicx}

\usepackage[title]{appendix}

\makeatletter 
\renewcommand\@biblabel[1]{#1.} 
\makeatother

\numberwithin{equation}{section}

\begin{document}


\baselineskip=17pt



\title{ Exponential scaling limit\\of the single-particle Anderson model
\\via adaptive feedback scaling  }

\author{ Victor Chulaevsky}


\date{}

\maketitle




\begin{abstract}
In this paper, we propose a reformulation of the bootstrap Multi-Scale Analysis (BMSA), developed
earlier by Germinet and Klein, as a single scaling algorithm,
to make explicit the fact that BMSA technique
implies an asymptotically
exponential decay of eigenfunctions (EFs) and of EF correlators (EFCs),
in the lattice Anderson models with diagonal disorder, viz. with an IID random
potential. We also show that the exponential scaling limit of the EFs and EFCs holds true
for a class of marginal distributions of the random potential with regularity lower than
H\"{o}lder continuity of any positive order.

\end{abstract}

\tmblue{
The version info: compared to v.1 of 9 March 2015, in the v.2 (21.07.2015) the scale-free threshold
has been slightly improved: $841^{−d} = 29^{−2d}$, as in [28], is replaced by $23^{−2d}$, owing to an
optimization of the geometrical argument in the proof of Lemma 1; the present text contains
additional figures; some intermediate technical parameters have been eliminated; all the remaining
parameters and the relations between them are listed in the tables (4.16)--(4.18) [(4.13)--(4.14) in v.3]; 
the proof of the
key Lemma 6 is rendered more detailed.
\vskip1mm
v.2 $\rightarrow$ v.3: Fig.6 has been added, and some numerical parameters has been rectified
in Fig.5.}


\section{Introduction and motivation}
\label{sec:intro}

We consider Anderson models with diagonal disorder in an integer lattice $\DZ^d$, $d\ge 1$.
Such models have been extensively studied over the last thirty years; the two principal tools
of the modern rigorous Anderson localization theory are the Multi-Scale Analysis (MSA)
and the Fractional Moment Method (FMM). In the framework of lattice systems (and more generally,
systems on graphs with sub-exponential growth of balls) the MSA proved to be more flexible;
in particular, it is less exigent to the regularity properties of the probability distribution generating
the local disorder -- in the simplest case, the single-site marginal distribution of the IID
(independent and identically distributed) values of the external random potential.
On the other hand, a considerable advantage of the FMM in the same class of models
is to provide exponential decay bounds for the (averaged) eigenfunction correlators (EFCs),
under the condition of H\"{o}lder continuity of the single-site marginal distribution. By comparison, the
original MSA scheme by Fr\"{o}hlich et al. \cite{FMSS85}, reformulated by von Dreifus and Klein \cite{DK89},
proved only a power-law decay of the key probabilistic estimates in finite volumes.
When the MSA was adapted to the proofs of strong dynamical localization (cf. \cite{GDB98,DS01,GK01}),
this resulted in power-law decay of EFCs.

Germinet and Klein \cite{GK01} significantly narrowed the gap between the EFC decay bounds provided by the
MSA and FMM. Specifically, using the bootstrap MSA, involving several interconnected
scaling analyses, they proved sub-exponen\-tial decay bounds with rate $L \mapsto \eu^{-L^\delta}$
for any $\delta\in(0,1)$. Their result is a finite-volume \emph{criterion} of exponential
spectral and sub-exponential dynamical localization:
if the latter occur for a random Hamiltonian of one of the forms considered in \cite{GK01},
then their scale-free condition is fulfilled for $L_0$ large enough.

Recently Klein and Nguyen
\cite{KN13a,KN13b} have adapted the BMSA to the multi-particle Anderson Hamiltonians.

In theoretical physics, the celebrated scaling theory, put forward by the "\emph{Gang of Four}"
(Abrahams, Anderson, Licciardello and Ramakrishnan, \cite{AALR79})
and further developing the Anderson localization theory \cite{A58},
predicted -- under certain assumptions
including also those sufficient for the MSA or FMM to apply -- that the functionals $F_L$
related to the quantum transport, first of all the conductance, for systems of large size $L$,
should admit a limiting behaviour in the double-logarithmic coordinate system.
While the existence of a.c. spectrum for systems on a periodic lattice
or in a Euclidean space remains an intriguing challenge for the mathematicians,
we show that in the parameter zone(s) where
various forms of localization can be established with the help of existing techniques,
the rate of decay $F(L)$ of eigenfunction correlators (EFCs) at large distances $L$ admits the limit
$$
 \lim_{L\to\infty} \frac{ \ln \ln F(L)}{ \ln L} =1.
$$
Below we will call such a behavior \emph{exponential scaling limit} (ESL).
Formally speaking, we obtain, as usual, only upper bounds, but the example of one-dimen\-sional systems
shows that decay faster than exponential should not be expected.

The main goal of the present paper is a transformation of the Germinet--Klein multi-stage
bootstrap MSA
procedure from Ref. \cite{GK01} into a single scaling algorithm, replacing several interconnected scaling analyses in the bootstrap method and establishing the ESL in the traditional Anderson model.

The motivation for the present work came from an observation, made in
Refs. \cite{C12a} (cf. \cite[Theorem 6]{C12a}), \cite{C14a} (see Theorem 8 in \cite{C14a}
and discussion after its proof),  and some earlier works,
that already in the von Dreifus--Klein method from Ref. \cite{DK89} there were some unexploited resources,
giving rise to ``self-improving'' estimates in the course of the induction on the length scales
$L_k, \, k\ge 0$,
following the recursion
$L_k = \lfloor L_{k-1}^\alpha \rfloor \sim L_0^{\alpha^k}$, $\alpha>1$. Specifically, it was observed that
the $k$-th induction step actually produces more decay of the GFs than required for
merely reproducing the desired decay rate
at the step $k+1$, and that  this excess can be put in a feedback loop, improving the master parameters of the scaling scheme.
The net result is the decay of the GFs (and ultimately, EFCs) faster than any power
law\footnote{This result holds true under a very weak regularity of the random potential,
just barely stronger than the conventional
log-H\"{o}lder continuity of the marginal distribution. See Assumption (W3) (Eqn. (2.15)) in \cite{C14a}.},
viz. $L\mapsto \eu^{ - a \ln^{1+c} L}$,
with $a,c>0$.

The benefits of such a feedback-based self-enhancement of the master scaling parameters
become much greater when the scales grow multiplicatively, as in the first stage of the BMSA:
$L_k = Y L_{k-1} = Y^k L_0$, with $Y\ge 2$. A fairly simple calculation shows that essentially
the same feedback loop as the one used in \cite{C12a,CS13,C14a} for the scales
$L_k \sim L_0^{\alpha^k}$, $k\ge 0$,
gives rise in this case to a fractional-exponential decay
$L\mapsto \eu^{ - L^{\delta}}$,
with some $\delta>0$.

Acting in the spirit of the bootstrap MSA, we implement a technically more complex
scaling procedure than the above mentioned ``simple feedback scaling",
aiming to render more explicit and constructive
the statement of the BMSA (cf. \cite{GK01})
that any (viz. arbitrarily close to $1$) value of the
exponent $\delta$ in the above formulae can be achieved for $L$ large enough. To this end, we
replace the first two stages of the BMSA (with fixed parameters) by an adaptive
feedback scaling algorithm. The latter makes the multiplicative growth factor $Y$,
figuring in the scaling relation
$L_k = Y L_{k-1}$, scale-dependent: $Y_k = \cY(k,L_k)$. In fact, the BMSA scheme includes another
important geometrical parameter -- an integer $S_k\in[1, Y_k)$; see Section \ref{sec:dominated}.

However, the "simple feedback scaling" -- with $Y_k$ and $S_k$ fixed -- may still be required during an
initial ``boost" stage, where the effects of localization are almost imperceptible,
particularly in the probabilistic estimates.
Since the scales grow with $k$ (viz. $L_k = Y^k L_0$), writing formally $Y_k = L_k^{\tau_k}$
results in a finite, initial sub-sequence $\{\tau_1, \ldots, \tau_{\fK-1}\}$, with some
$\fK$ depending upon the model parameters, which is actually decreasing. (As such, the values
$\{\tau_1, \ldots, \tau_{\fK-1}\}$ are simply unused.)
It is only later, for $k > \fK$, that we fix $\tau_k = \tau>0$, thus effectively switching
to the super-exponential growth $L_k \sim C L_\fK^{(1+\tau)^k}$. Of course, depending on the reader's
personal point of view, the presence of this switching point may be considered as a
form of the Germinet--Klein multi-stage technique.

Taking into account the abundance of various technical parameters in our scheme, we keep $\tau_k$
fixed for the rest of the scaling procedure. However, the algorithm's efficiency can be further
improved by making $\tau_k$ also $k$-dependent (and growing).
This may prove useful in a numerical implementation of the adaptive scaling
algorithm, as well as in specific models (including the multi-particle models with slowly decaying
interaction).
We show that the "gap" between the genuine exponential decay (i.e., the value $\delta =1$)
and the exponent $\delta_k$ achieved at the $k$-th step, decays at least exponentially fast
in $k$. In a way, this
provides a rigorous complement to the predictions of the physical scaling theory
on the convergence to the ESL, at least in the parameter zone(s) where localization can be proved
with the existing scaling methods.

The core of the renormalization group-style analysis, the proof of the key Lemma \ref{lem:scaling.prob},
is not very long, but tedious, and the number of intermediate scaling exponents and other parameters
is quite disturbing, so the reader may naturally wonder if the entire inductive algorithm could be made
significantly simpler and more transparent. The answer is affirmative, but this would require
a certain concession: as show our preliminary calculations, it suffices to start with
the initial probability $p_0$ (cf. \eqref{eq:cond.p0.tables}) which is not scale-free but polynomial in $L_0$,
i.e., $p_0 \le L_0^{-b_0}$ with a value of $b_0>0$ which is, actually, not excessively large.
This would eliminate the need for
the additional "boost" phase in the scaling algorithm, where the decay of the probabilities
of unwanted events is barely perceptible (and virtually useless for applications
to physically
realistic\footnote{To have an idea of typical situations where Anderson localization theory
proves valuable in modern technology, recall that a typical CPU is today a square film with
about $10^7\div 10^8 $ atoms alongside ($\sim 1 {\rm cm}$) and just a few atoms ($\sim 20$ nm) across it.}
models), and also render the whole procedure more straightforward.
In the author's opinion, this would be indeed only a small concession, for two reasons:
\begin{itemize}
  \item the main role of the scale-free Germinet--Klein's criterion is to provide a mechanism
  of transforming weak probabilistic bounds into much stronger ones, and this task
  is entrusted to the adaptive  feedback that we propose;

  \item more importantly, the actual verification of the initial conditions for the onset of Anderson
  localization is usually based on one of the two methods: using the Lifshitz tails phenomenon
  (providing de facto very strong, fractional-exponential bounds), or deriving the initial bounds from
  the strong disorder assumption, which also provides strong \emph{scale-dependent} probabilistic bounds.
\end{itemize}

Still, we keep in the present paper the scale-free localization criterion to show that
the Germinet-Klein BMSA extends to the result on exponential scaling limit without any additional hypotheses.

Speaking of the consecutive phases (analyses) of the BMSA, it is to be pointed out
that we do not perform the last stage where a genuine exponential decay of the Green functions
is established in cubes of size $L_k$ with probability $\sim 1 -\eu^{-L_k^{\delta_k}}$,
where $\delta_k = \delta$ is made arbitrarily close to $1$ by the results of \cite{GK01};
one would expect $\delta_k \nearrow 1$ in the framework of the present paper. We do not
analyze the behaviour of such probabilities related to the
\emph{exponential} decay of the GFs in finite volume.
As was already said, this paper focuses on the exponential scaling \emph{limit} -- for the Green functions, eigenfunctions and eigenfunction correlators. The actual road map is as follows:
$GFs \rightsquigarrow EFCs \rightsquigarrow EFs$, so the decay rate of the EFs is shaped by that of the
EFCs. Naturally, one can switch at any moment
from the analysis of the "almost exponential" decay to that
of the exponential one, by simply following the Germinet-Klein approach, but our main goal
is the construction of a \emph{single}  algorithm which takes care of all exponents
$\delta$ close to $1$. Undoubtedly, there can be various further developments
of the BMSA technology from \cite{GK01}.

Finally, we show that the proposed adaptive feedback scaling technique allows for a
lower regularity of the  marginal distribution
of the IID random potential than H\"{o}lder continuity of any positive order.
In the realm of the FMM proofs of localization, it is well-known that  absolute continuity of the
marginal distribution can be safely and easily relaxed
to H\"{o}lder continuity of any positive order $\fbH$ (cf. \cite{AM93})
Yet, the MSA in general is renowned for its higher
tolerance to a lower
regularity of the probability distribution of the disorder. So, while the question on the lowest regularity
compatible with the FMM approach to the \emph{exponential} strong dynamical localization remains open,
our results evidence that H\"{o}lder continuity is not required for the exponential scaling \emph{limit}
of the EF correlators.

As was said, strong dynamical localization at \emph{some} fractional-exponential rate
$\delta\in(0,1)$ actually follows from
the initial, weak hypotheses through a simpler scaling procedure,
under the assumption of H\"{o}lder continuity of the
marginal PDF of the random potential.

We would like to make one last comment, hopefully providing an answer to the readers,
familiar with the FMM approach, who may wonder: does one really need an alternative method, such as
(a version of) the MSA, to prove "almost" exponential localization in the models with
sufficiently regular probability distribution of the random potential, where the FMM
proves a genuine exponential decay of EFs and of EFCs? Well, there are several
elements that have to be taken into account.

\begin{itemize}
  \item Firstly, the MSA or its variants have been
successfully applied to the models with deterministic disorder where the dependence between
the values of the potential at distinct points is so strong that no decoupling
inequalities or similar techniques, which have been the cornerstone of the Aizenman--Molchanov
method since the pioneering work \cite{AM93}, seem to apply.

  \item Secondly, the Fractional Moments Method
ceases to be a "mono-scale" technology in the area of multi-particle Anderson localization.
Indeed, both the paper by Aizenman and Warzel \cite{AW09} and a recent work by Fauser and Warzel
\cite{FW15} are based on a \emph{scale induction} for the fractional moments of the Green functions.
This is unrelated to the regularity of the marginal distribution.

  \item
Thirdly, in the $N$-particle Anderson models with interaction decaying slower than exponentially,
the only known proofs of genuine exponential decay of the EFs (cf. \cite {CS15a,C14e})
employ an important particularity of the MSA: the analyses of the EFs and of their correlators
can be carried out independently, so one can prove exponential decay of the EFs even
in some situations where the EFCs decay sub-exponentially. In the framework of the FMM,
the analysis of the EFs is subordinate to that of the EFCs. Again, this issue arises
even in the most regular models of the disorder.

\end{itemize}

Summarizing, mathematical methods are not always ordered on a linear scale,
and new challenging models may test and contest the relations between them.

\subsection{The model}

We focus on the case where the configuration space of a quantum particle in an external
random potential is the lattice $\DZ^d$, $d\ge 1$, and consider the random Hamiltonian $H(\om)$
of the form
\be\label{eq:def.H}
\big(H \psi\big)(x) = \sum_{|y-x|_1=1} \big( \psi(x) - \psi(y) \big) + V(x;\om) \psi(x),
\ee
where $V: \DZ^d\times\Om\to \DR$ is an IID random field relative to some probability
space $(\Om,\fF,\DP)$, and $|x|_1 := |x_1| + \cdots + |x_d|$ for $x=(x_1, \ldots, x_d)$.
Until Section \ref{sec:lower.reg}, we assume that
marginal probability distribution function (PDF) $F_V$,
of the random field $V$,
$$
F_V(t) := \pr{ V(0;\om) \le t}, \;\; t\in\DR,
$$
is H\"{o}lder-continuous of some order $\fbH\in(0,1)$. In Section \ref{sec:lower.reg} we show
that the assumption of H\"{o}lder-continuity can be slightly relaxed (cf. Eqn. \eqref{eq:weak.cond.sV}).

The second-order lattice Laplacian in \eqref{eq:def.H}
can be easily replaced by any (self-adjoint) finite-difference
Hamiltonian of arbitrary finite order, without any significant modification of our algorithm. Indeed, we replace
the form of the Geometric Resolvent Inequality most often employed in the MSA of lattice models,
with its variant traditional for the MSA in continuous systems (in $\DR^d$). It is based on a simple
commutator relation, so that the range (order) of a finite-difference kinetic energy operator
becomes irrelevant (and unused in the intermediate calculations),
provided the initial length scale $L_0$ is large enough. For clarity, we work
only with the standard lattice Laplacian.

\subsection{The main assumption}

The principal assumption on the parameters of the model at hand is the probabilistic inequality
\eqref{eq:def.prob.p_0} which we formulate in Sect.~\ref{ssec:AFS.tech}, for it requires several
definitions given there.

\subsection{Structure of the paper}
\label{ssec:structure}

\begin{itemize}
  \item[$\bullet$] The principal objects and notations are introduced in Section \ref{sec:basic}.

  \item[$\bullet$] In Section \ref{sec:dominated}, we present the main analytic tool of the scaling analysis -- the
  Geometric Resolvent Inequality  (GRI), and formulate the main result of the paper,
  Theorem \ref{thm:Main.ESL}, providing a scale-free criterion for the exponential scaling limit of the
  Green functions. The exposition is closer
  to the form of the GRI used in the continuous systems than to the one traditionally used in
  the lattice models, starting from the pioneering papers \cite{FS83,FMSS85,Dr87,Sp88,DK89}.
  This is required   for the geometrical optimizations \`{a} la Germinet--Klein \cite{GK01}.

  \item[$\bullet$] The core of the paper is Section \ref{sec:ASFS}, and the staple there is Lemma \ref{lem:scaling.prob}
  concluding the fixed-energy MSA (FEMSA).

  \item[$\bullet$] The derivation of the exponential scaling limit from the results of Section \ref{sec:ASFS}
  is given in Section \ref{sec:ESL}.

  \item[$\bullet$] Section \ref{sec:FEMSA.to.DL} is devoted to a "soft" derivation
  of the variable-energy MSA (VEMSA) and of strong dynamical
  localization from FEMSA carried out in Section \ref{sec:ASFS}.

  \item[$\bullet$] In Section \ref{sec:lower.reg}, we show that the assumption of H\"{o}lder-continuity
  of the  marginal probability distribution of the random potential an be relaxed. To the best
  of the author's knowledge, this result is new.

  \item[$\bullet$] A number of definitions of various technical parameters and relations between them
  are listed in Sect.~\ref{ssec:tables}, in the tables \eqref{eq:table2a}--\eqref{eq:table2b}.
  The reader may find it helpful to have them printed on a separate page
  when checking the proofs.
\end{itemize}

In theoretical physics, a sufficiently fast  decay of the Green functions
away from the diagonal is usually considered as one of equivalent signatures
of Anderson localization. Speaking mathematically, this is a higher-dimensional
analog of positivity of Lyapunov exponents in one-dimensional (or quasi-one-dimensional)
systems. While it is known that, in general, this analog does not necessarily imply spectral localization,
first, it has been shown long ago by Martinelli and Scoppola \cite{MS85}
that it rules out a.c. spectrum with probability one, and secondly, it has been observed
that the s.c. spectrum occurs in systems with some strong ``degeneracies'' in the
probability distribution of the ergodic (not necessarily IID or weakly correlated)
potential. Under reasonable assumptions on regularity of the IID random potential, fast
decay of the GFs implies indeed spectral and strong dynamical localization, and the role
of Section \ref{sec:FEMSA.to.DL} is to summarize the progress achieved in this direction, and
to show in a fairly simple way that the fixed-energy analysis is the heart of the localization
analysis of the conventional lattice Anderson model.

The continuous systems are not considered in the present paper,
since the analysis of unbounded (differential) random operators would  require an additional
technical discussion
pertaining to the domains, self-adjointness, etc. But as was already said, here we focus mainly on
the scaling algorithm that could be applied, essentially in the same way, both to the discrete
and continuous systems.
\vskip1mm

\section{Basic geometric objects and  notations}
\label{sec:basic}

Following essentially Ref. \cite{GK01} (where the Anderson-type models in a continuous space $\DR^d$
were considered), we work with a hierarchical collection of lattice cubes, with specific centers
and positive integer side lengths $L_k$.
For our purposes, it is more convenient to start with the cardinalities of the cubes and those of their
one-dimensional projections: we fix odd positive integers $Y>1$, $\ell_0$ and set
$$
L_k = Y^k \cdot 3 \ell_0 = 3 \cdot Y^k \ell_0 =: 3 \ell_k.
$$
(At some moment in the proofs, the scaling factor $Y$ becomes variable: $Y = Y_k$.)
Next, we consider the lattice cubes with coordinate projections of cardinality $L_k$:
$$
\ball_{L_k}(x) := \myset{ y\in\DZ^d:\; |y-x|\le \frac{L_k}{2} }, \;\; |x| := \max_i |x_i| .
$$
Since $L_k = 3 Y^k \ell_0$ is odd, the upper bound in the above definition of the cube
$\ball_{L_k}(x)$ could have been replaced with $(L_k-1)/2$, resulting in the same lattice subset.
However, having in mind the canonical embedding $\DZ^d \hookrightarrow\DR^d$, the above definition
looks more natural when transformed as follows: with $y\in\DZ^d\hookrightarrow\DR^d$,
$$
\ball_{L_k}(x) := \myset{ y\in\DZ^d\hookrightarrow\DR^d:\; |y-x|\le \frac{L_k}{2} } ,
$$
so that the "fictitious" radius of the ball is precisely $L_k/2$.

Sometimes it is more convenient to refer to the spherical layers and balls relative to the max-distance,
with a clearly identified integer \emph{radius}:
\be\label{eq:def.cL.Lam}
\bal
\cL_r(u) &= \myset{x\in\DZ^d:\, |x-u| = r } ,
\\
\Lam_r(u) &= \myset{x\in\DZ^d:\, |x-u| \le r } \equiv \ball_{2r+1}(u)\,.
\eal
\ee
Notice that one has
$\ball_{L_k}(u) = \Lam_{\frac{L_k - 1}{2}}(u)$.

The cube $\ball_{L_k}(u)$
is partitioned into $3^d$ adjacent cubes called $k$-cells,
\be\label{eq:Lamk.Ballk}
 \scorek(c) :=  \ball_{\ell_k}( c ) = \Lam_{ \frac{\ell_k - 1}{ 2 }}(c)
\ee
(recall: $3\ell_k = L_k$) with centers $c$ in the sub-lattice
$(3\DZ)^d$.
\begin{itemize}
  \item[$\bullet$] The central cell $\scorek(u)$ of a cube $\ball_{L_k}(u)$ will be called the \emph{core} of $\ball_{L_k}(u)$;

  \item[$\bullet$] the complementary annulus, formed by the remaining $3^d$-1 cells of $\ball_{L_k}(u)$,
  will be called the \emph{shell} of $\ball_{L_k}(u)$.
\end{itemize}

Given any length scale $L_k = Y^k L_0$, we shall always work with the family of $L_k$-cubes
whose cells form the uniquely defined partition of $\DZ^d$ including the cube centered at the origin,
$\ball_{\ell_k}(0)$; these cores, as well as their centers, will be called \emph{admissible}
at the scale $L_k$. The centers of the admissible $\ell_k$-cores form a sub-lattice of $\DZ^d$
denoted by $\cCk$. Sometimes we use notation $\lr{c,c'}$,  meaning
that $c, c'\in\cCk$ are two nearest neighbors (in $\cCk$) relative to the max-distance: $|c - c'|=\ell_k$.
By a slight abuse of notations, we will write, e.g.,  $\sum_{\lr{c,c'}\in\cCk}$
instead of $\sum_{\lr{c,c'}\in\cCk \times \cCk}$
Each point $c\in\cCk$ has $3^d - 1$ nearest neighbors (within $\cC^k$).

\begin{figure}
\begin{tabular}{c}
%
%
%
\begin{tikzpicture}
\begin{scope}[scale=0.12]
\clip (-35,-9.5) rectangle ++(45.0, 21.0);

\foreach \i in {-1.5, 1.5}
{
  \draw[color=white, line width = 2] (\i, -4.5) -- (\i, 4.5);
  \draw[color=white, line width = 2] (-4.5, \i) -- (4.5, \i);
}

\fill[color=black!50!white!50] (-4.5, -4.5) rectangle ++(9, 9);

\draw[color=white] (-1.6, -4.8) -- (-1.6, 4.8);
\draw[color=white] (1.6,  -4.8) -- (1.6,  4.8);

\draw[color=white] (-4.8, -1.6) -- (4.8, -1.6);
\draw[color=white] (-4.8,  1.6) -- (4.8,  1.6);

\fill[color=black!80!white!50] (-1.5, -1.5) rectangle ++(3,3);

\foreach \nx in {-4, -3, ..., 4}
{
  \foreach \ny in {-4, -3, ..., 4}
  {
    \fill[color=black] (\nx, \ny) circle (0.1);
  }
}

\foreach \i in {-9, -6, ..., 9}
{
\foreach \j in {-9, -6, ..., 9}
  \fill[color=black] (\i, \j) circle (0.3);
}

\end{scope}
\end{tikzpicture}
%

\end{tabular}

\caption{  \footnotesize\emph{Cubes and cells}.
\footnotesize
Here $d=2$, $L_0=9$. $3^2-1$ cells forming the shell of a cube $\ball_9(\cdot)$ is shown in gray color, 
and the core in dark gray.
}
\end{figure}
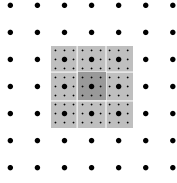%

See Fig.~1 where
\begin{itemize}
  \item[$\bullet$] an admissible square  of size $L=9$ (thus with $9$ vertices along each side) is shown in gray color;
  it is partitioned into $3^d = 3^2$ congruent cells separated visually by thin white lines; the admissibility
  means that the periodic sublattice of the cell centers (large black dots) includes the origin $0\in\DZ^d$;

  \item[$\bullet$] the \emph{core}, i.e., the central cell, is shown in a darker shade of gray than the shell;

  \item[$\bullet$] each cell is composed -- in this example -- of $3^2$ points.
\end{itemize}

\noindent
The larger dots on Fig.~1
represent the centers of the cells of size $\ell_0=L_0/3=3$ admissible in the geometrical constructions
referring to the cubes of such size. In this case, the minimal spacing between the centers of admissible cores
equals $3$. Considering
$L_0=9$, we have the spacing $\ell_0= L_0/3$. The admissible cells of a given size form a partition of $\DZ^d$,
and we denote by $\scorek(x)$ the unique admissible cell of size $\ell_k=L_k/3$,
containing a given point $x$; naturally, $\scorek(x)=\scorek(y)$ for all $x,y$
from the same admissible cell, so there is no conflict with the previously introduced
notation $\scorek(c)$ where $c$ was the cell's center.

It will be convenient to endow the set of the admissible cell centers $c\in\cC^k$ in $\ball_{L_{k+1}}(u)$
with the natural graph structure, with edges formed by the pairs of nearest neighbors $c, c'$
with respect to the max-distance, i.e., those with $|c-c'| = \ell_k= L_k/3$. Such a graph
$\cB_{k+1}$ will be called the \emph{skeleton graph} of $\ball_{L_{k+1}}$. Skeleton graphs
will be used in Appendix \ref{app:domin.decay}.

The main tool for the analysis of the Green functions in such cubes often
is the Geometric Resolvent Inequality (GRI). In its basic form, used in \cite{Dr87,Sp88,DK89}
and in numerous subsequent works,
a single application of the GRI moves one from the center of a given cube $\ball_L(x)$ to (any) point $y$
of the exterior boundary
$\pt^+ \ball_L(x) := \big\{z:\, \rd\big(z, \ball_L(x)\big) = 1) \big\}$. Here
$\rd(\cdot\,, \cdot)$ stands for the graph-distance in the lattice $\DZ^d$, with edges formed by the
nearest neighbors
in the norm $|\cdot|_1$. The notion of the exterior boundary is relative to an ambient set
$\Lam \supset \ball_L(x)$
(a subgraph of $\DZ^d$), when the analysis is carried out in a proper subset $\Lam$ of the lattice.
As was said in Section \ref{ssec:structure}, we employ another version of the GRI
typical for the applications to the continuous Anderson models (cf., e.g., \cite{GK01}).

Given a finite subset $\Lam\subset \DZ^d$, we introduce the local Hamiltonian
$H_\Lam := \one_\Lam H \one_\Lam \upharpoonright \ell^{2}(\Lam)$, acting in the
finite-dimensional Hilbert space $\ell^{2}(\Lam)$ canonically injected into $\ell^{2}(\DZ^d)$.
$H_\Lam$ is self-adjoint; it is often considered as the restriction of $H$
to the subset $\Lam$ with Dirichlet boundary conditions outside $\Lam$, but
the terminology here varies from one source to another (cf., e.g., \cite[Sect.~5.2]{Kir08a}).
We introduce the interior boundary $\pt^-\Lam$ and exterior boundary $\pt^+\Lam$  of
the set $\Lam$ by
$$
\bal
\pt^- \Lam &:= \{x\in\Lam:\, \dist(x, \DZ^d\setminus \Lam)=1\} ~,
\;\;\;
\pt^+ \Lam := \pt^- \big(\DZ^d\setminus \Lam \big) ~.
\eal
$$
For brevity, we often use notations like $\Sigma_\Lam$ for the spectrum of
$H_\Lam$, i.e., the set of its eigenvalues, counting multiplicity. In the case
where $\Lam  = \ball_L(u)$, we also write $\Sigma_{u,L}$. This definition will be
recalled where necessary.

For the derivation of the GRI, it is convenient to use the language of the
balls $\Lam_r$ of explicitly specified radius $r$, rather than the balls $\ball_L$, with
$L$ directly related to the cardinality of each coordinate projection.
Let $\Lam = \Lam_{R_k}(c_x) \subset \Lam'$, for some finite $\Lam'\subset\DZ^d$
with $\rd(\pt^+ \Lam, \pt^- \Lam') \ge 2$,
$\phi = \one_{\Lam_{R_k - 1}}$, and $\rPhi$ be the operator of multiplication by $\phi$.
Note that for any $u\in\Lam_{R_{k-1}}(c_x)$,
i.e., any $\one_u \in \Ran \rPhi$,
we have the identities $\one_{\Lam'} \one_u  = \one_u  = \one_{\Lam} \one_u $,
and
$$
\supp \big( H \one_u  \big)\subset \Lam_{R_k} \equiv \Lam ~,
$$
since $H$ is a finite-difference operator of order $2$.
Therefore,
$
 H \one_u =  \one_{\Lam} H \one_u,
$
and similarly, using $\one_{\Lam'}\one_{\Lam} = \one_{\Lam}$,
$$
\bal
H_{\Lam} \one_u &=
 \one_\Lam H \one_\Lam  \one_u = \one_{\Lam} H \one_u ~,
\\
H_{\Lam'} \one_u &=
 \one_{\Lam'} H \one_{\Lam'}  \one_u = \one_\Lam H \one_u ~,
\eal
$$
so for any basis vector $\one_u \in \Ran \rPhi$,
$$
H_{\Lam'} \one_u = H_{\Lam} \one_u ~.
$$
As a result, one has the operator identity
$H_\Lam \rPhi = H_{\Lam'} \rPhi$, thus for any $E\in\DR$,
\be\label{eq:commut}
\bal
  (H_\Lam - E) \rPhi &= (H_{\Lam'} - E)\rPhi
\\
   & = \rPhi (H_{\Lam'} - E) - [\rPhi, (H_{\Lam'} - E) ]~.
\eal
\ee
Below, the energy $E$ will be fixed and omitted from notation in the resolvents
$G_{\Lam'} = (H_{\Lam'} -E)^{-1}$, $G_{\Lam} = (H_{\Lam} -E)^{-1}$.
Denoting
\be\label{eq:def.W}
W = [ \rPhi,  (H_{\Lam'} - E) ] \equiv [ \rPhi,  \Delta_{\Lam'} ]
\ee
(note that $[\rPhi,V-E]=0$) and multiplying the identity stemming from \eqref{eq:commut},
$$
\rPhi (H_{\Lam'} - E) = (H_\Lam - E) \rPhi + W~,
$$
by $G_{\Lam}$ on the left and by $G_{\Lam'}$ on the right,
we obtain the identity
$$
 G_{\Lam} \rPhi = \rPhi G_{\Lam'}  + G_{\Lam} W G_{\Lam'} \, .
$$
The RHS representation of $W$ in \eqref{eq:def.W}
implies
\be\label{eq:norm.W}
\|W\|\le 2 \|\rPhi\|\, \|\Delta_{\Lam'}\|\le 8d .
\ee
Due to the relations \eqref{eq:def.cL.Lam} and \eqref{eq:Lamk.Ballk}
we have $\ball_{L_k}(c_x) = \Lam_{R_k}(c_x)$, with $R_k = (L_k-1)/2$.
Let $\rGamk_{x}$ be
the boundary annulus of width $2$ of $\ball_{L_k}(c_x)$
(cf. Fig. 2):
\be\label{eq:def.Gamk}
\rGamk_x \equiv \rGamk_{c_x}:= \Lam_{R_k}(c_x) \setminus \Lam_{R_k -2}(c_x),
\ee
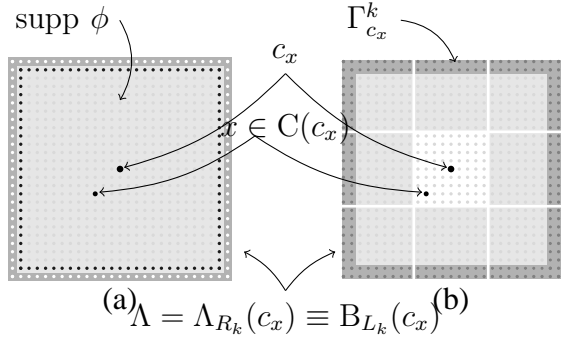
\begin{figure}
\begin{tabular}{c}
%
\begin{tikzpicture}
\begin{scope}[scale=0.11]
\clip (-70,-22) rectangle ++(90.0, 45.0);

\fill[color=black!30!white] (-13.5-40,-13.5) rectangle ++(27.0, 27.0);
\fill[color=gray!20!white] (-12.5-40,-12.5) rectangle ++(25.0, 25.0);

\foreach \i in {-12, -11, ..., 12}
{
  \foreach \j in {-12, -11, ..., 12}
  {
    \fill[color=gray!30!white]  (\i-40, \j) circle(0.2);
  }
}

\foreach \i in {-13, -12, ..., 13}
{
\fill[color=black!0!white]  (\i-40, +13) circle(0.2);
\fill[color=black!0!white]  (\i-40, -13) circle(0.2);

\fill[color=black!0!white]  (-13-40, \i) circle(0.2);
\fill[color=black!0!white]  (+13-40, \i) circle(0.2);
}

\foreach \i in {-12, -11, ..., 12}
{
\fill[color=black!90!white]  (\i-40, +12) circle(0.2);
\fill[color=black!90!white]  (\i-40, -12) circle(0.2);

\fill[color=black!90!white]  (-12-40, \i) circle(0.2);
\fill[color=black!90!white]  (+12-40, \i) circle(0.2);
}

\node (fig_a) at (0 - 40, -16) {(a)};

\node (supp_phi) at (-7 - 40, 18) {$\supp\, \phi$};
\node (phi_arr) at (0 - 40, 7) {};
\draw[->, bend left = 30]  (supp_phi.east) to (phi_arr.north);


\fill[color=black!30!white] (-13.2,-13.2) rectangle ++(26.4, 26.4);
\fill[color=gray!20!white] (-11.5,-11.5) rectangle ++(23.0, 23.0);
\fill[color=white] (-4.5,-4.5) rectangle ++(9.0, 9.0);

\foreach \i in {-12, -11, ..., 12}
{
  \foreach \j in {-12, -11, ..., 12}
  {
    \fill[color=gray!30!white]  (\i, \j) circle(0.2);
  }
}

\foreach \i in {-12, -11, ..., 12}
{
\fill[color=black!50!white]  (\i, +12) circle(0.2);
\fill[color=black!50!white]  (\i, -12) circle(0.2);

\fill[color=black!50!white]  (-12, \i) circle(0.2);
\fill[color=black!50!white]  (+12, \i) circle(0.2);
}

\foreach \i in {-13, -12, ..., 13}
{
\fill[color=black!50!white]  (\i, +13) circle(0.2);
\fill[color=black!50!white]  (\i, -13) circle(0.2);

\fill[color=black!50!white]  (-13, \i) circle(0.2);
\fill[color=black!50!white]  (+13, \i) circle(0.2);
}

\foreach \i in {-4.5, 4.5}
{
  \draw[color = white, line width = 1] (\i, -13.4) -- (\i, 13.4);
  \draw[color = white, line width = 1] (-13.4, \i) -- (13.4, \i);
}

\node (fig_b) at (0, -16) {(b)};

\node (supp_Gam) at (-10, 18) {$\rGam^k_{c_x}$};
\node (Gam_arr) at (0.5, 11.6) {};
\draw[->, bend left = 30, color=black]  (supp_Gam.east) to (Gam_arr.north);

\fill[color=black] (0,0) circle (0.4);
\fill[color=black] (-40,0) circle (0.4);
\node (c_x) at (-20, 14) {$c_x$};
\node (x) at (-20, 5) {$x \in \rC(c_x)$};

\draw[->,bend right=15] (c_x.south) to (-0.5,0.2);
\draw[->,bend left=15,color=black]  (c_x.south) to (-40+0.5,0.2);

\fill[color=black] (-3,-3) circle (0.3);
\fill[color=black] (-40-3,-3) circle (0.3);
\draw[->,bend right=15] (-23.7, 4.0) to (-3-0.5, -3+0.2);
\draw[->,bend left=15] (-23.7, 4) to (-40-3+0.5, -3+0.2);

\node (Lam) at (-20, -18) {$\Lam = \Lam_{R_k}(c_x) \equiv \ball_{L_k}(c_x)$};

\draw[->,bend right=15,color=black]  (Lam.north) to (-25, -10);
\draw[->,bend left=15,color=black]  (Lam.north) to (-14, -10);

\end{scope}
\end{tikzpicture}
%
\end{tabular}

\caption{
\footnotesize
Example for the GRI \eqref{eq:GRI.1}.
Black dots on Fig.2 (a) represent the boundary of $\supp\,\phi$, and the white
dots -- the boundary of $\DZ^d \setminus \supp\, \phi$. Their union is the set
$\rGamk_x \equiv \rGamk_{c_x}$, shown as the dark gray annulus on Fig. 2 (b). The commutator $W$
annulates any function supported by the complement $\DZ^d\setminus \rGam^k_{c_x}$.}
\end{figure}%
and $\bGamk_x \equiv \bGamk_{c_x}$ (boldface notation) the operator of multiplication by $\one_{\rGamk_{x}}$.
More generally, in the case where $R=(L-1)/2$, $y\in\DZ^d$, $\ball = \ball_L(y)$, denote
\be\label{eq:def.Gam.ball}
\rGamk_\ball := \Lam_{R}(y) \setminus \Lam_{R -2}(y),
\ee
and let $\bGamk_\ball$ be the operator of multiplication by
$\one_{\rGamk_\ball}$. Introduce also a shortcut for the minimal set of centers
of $L_k$-admissible cells covering a given subset $A\subset\DZ^d$:
\be\label{eq:def.cCk.A}
\cCk(A) = \{c\in\cCk:\, \cell_k(c)\cap A \ne \varnothing\}.
\ee
Observe that the Laplacian $\Delta$, being the canonical graph Laplacian (on $\DZ^d$),
annulates the constant functions, hence
$$
\supp \big( \Delta \phi \big) \subset \rGamk_{c_x}
$$
(in fact, $\supp \big( \Delta \phi \big) = \rGamk_{c_x}$).
Indeed, for any $z\in \Lam_{R_k}(x) \setminus \rGamk_{c_x}$, the function $\phi \equiv \one_{\Lam_{R_k - 1}}$
takes the constant value $1$ on the $1$-neighborhood of $z$, thus $(\Delta \phi)(z)=0$.
Similarly, $\phi$ vanishes on the $1$-neighborhood of any point $z \in\DZ^d \setminus \Lam_{R_k}(c_x)$,
hence $(\Delta \phi)(z)=0$. In other words, $\supp \big( \Delta \phi \big)$ is covered by
the union of the boundary of $\supp \phi = \one_{\Lam_{R_k-1}}(c_x)$ and of the boundary
of $\supp(1 -\phi)$.

It follows that the commutator $W =[ \rPhi,  (H_{\Lam'} - E) ] = [ \rPhi,  (\Delta_{\Lam'} - E) ]$
satisfies the operator identity
$W = \bGamk_x W \bGamk_x$,
so for any subset
$A\subset \Lam' \setminus \Lam_{R_k}(u)$, one has
\be\label{eq:GRI.1.first}
\bal
 \one_A G_{\Lam'} \chik_{c_x} &
 =  \one_A \rPhi G_{\Lam} \chik_u  + \one_A  G_{\Lam'} W G_{\Lam} \chik_{c_x}
\\
 & =  \big( \one_A  G_{\Lam'} \bGamk_x \big) \, W  \, \big( \bGamk_x G_{\Lam} \chik_{c_x}  \big) ,
\eal
\ee
and we come to the following form of the Geometric Resolvent Inequality:
\be\label{eq:GRI.2}
  \big\| \one_A G_{\Lam'} \chik_{c_x} \big\|
     \le \big\|  W \big\| \,\cdot\big\| \one_A  G_{\Lam'} \bGamk_x \big\|  \cdot \big\|  \bGamk_x G_{\Lam} \chik_{c_x} \big\| .
\ee
Introduce a slightly abusive but convenient notation, recalling that we are going
to use a sequence of length scales following the recursion $L_{k} = Y_{k} L_{k-1}$ :
\be\label{eq:def.dnorm}
\dnorm{ G_{\ball_{L_k}(x)}} :=
   \CWk \big\| \one_{\bGamk_x} G_{\ball_{L_k}(x)}(E) \chik_{c_x} \big\| ,
   \;\; \CWk :=  Y_k^d\| W \| ~.
\ee
Here $\curlywedge$ symbolizes the decay from the center to the boundary of a cube.
A more accurate (but cumbersome) notation would include the dependence of the
symbol $\curlywedge$ upon the cube $\ball$.
For brevity, let $\ball' = \ball_{L_{k+1}}(u)$, $\ball = \ball_{L_{k}}(x)$.
With $A = \rGamkone_u$ (the set $A$ appears in \eqref{eq:GRI.1.first}--\eqref{eq:GRI.2}),  we
infer from \eqref{eq:GRI.2} and $\| W \| = Y^{-d} \CWk $
\be\label{eq:GRI.1}
\bal
\big\| \bGamkone_u G_{\ball'} \chik_{c_x} \big\|
     &\le Y^{-d} \CWk  \, \cdot\big\| \bGamkone_u  G_{\ball'} \bGamk_x \big\|
     \cdot \big\|  \bGamk_x G_{\ball} \chik_{c_x} \big\|
\\
  & \le  Y^{-d}  \CWk
  \cdot \big\| \bGamk_x  G_{\ball} \chik_{c_x} \big\|
  \sum_{ c: |c-c_x| = \ell_k }
      \big\| \bGamkone_u  G_{\ball'} \chik_{c} \big\|
\\
  & \le  \dnorm{ G_{\ball} } \max_{ \lr{c, c_x}\in \cCk }
      \big\| \bGamkone_u  G_{\ball' } \chik_{c_x} \big\| ~,
\eal
\ee
as $\card\{c:\, \lr{c, c_x}\in \cCk \} < Y^{d}$.
This bound is useful when $\dnorm{ G_{\ball} }$ is small.

By self-adjointness of our Hamiltonians, we also have
\be\label{eq:bound.norm.dist.Sigma}
\bal
 \dnorm{ G_{\ball} } = \CWk \big\| \bGam_\ball G_{\ball} \chik_{c_x} \big\|
&\le  \CWk \big\| \bGam_\ball \big\| \, \big\| \chik_{c_x} \big\|
 \;  \left( \dist(E, \Sigma( H_\ball) \right)^{-1}
\\
&\le   \CWk \left( \dist(E, \Sigma( H_\ball) \right)^{-1} ,
\eal
\ee
yielding an \emph{a priori} bound  useful in the case where
$\dnorm{ G_{\ball} }$ is not small:
\be\label{eq:bound.GRI.dist.Sigma}
\bal
\big\| \bGam^{k+1}_{u} G_{\ball'} \chik_{c_x} \big\|
&\le \frac{\CWk }{\dist\big(E, \Sigma( H_{\ball})\big) }
    \max_{\lr{c, c_x}\in \cCk} \big\| \bGam^{k+1}_{u} G_{\ball'} \chik_c \big\| ~.
\eal
\ee
More generally, in the case where (cf. Fig. 3)
\be\label{eq:general.ball}
x \in \ball = \ball_{L}(w)\subset \ball_{L_{k+1} }(u),
\; \dist\big( \ball_L(w), \rGam^{k+1}_u \big) >0,
\ee
with an arbitrary $L < L_{k+1}$  compatible with \eqref{eq:general.ball},
the boundary belt $\rGam_\ball$ figuring in \eqref{eq:bound.norm.dist.Sigma}
is covered by at most (in fact, less than) $Y^d$ cells $\scorek(c)$, $c\in\cC^k$.
%
%
Therefore, in such a general case,
we still have
\be\label{eq:bound.GRI.dist.Sigma.general}
\bal
\big\| \bGam^{k+1}_{u} G_{\ball'} \chik_{c_x} \big\|
&\le \frac{\CWk }{\dist\big(E, \Sigma( H_{\ball})\big) }
    \max_{c\in \cCk(\rGamk_\ball)}
    \big\| \bGam^{k+1}_{u} G_{\ball'} \chik_c \big\| ~.
\eal
\ee

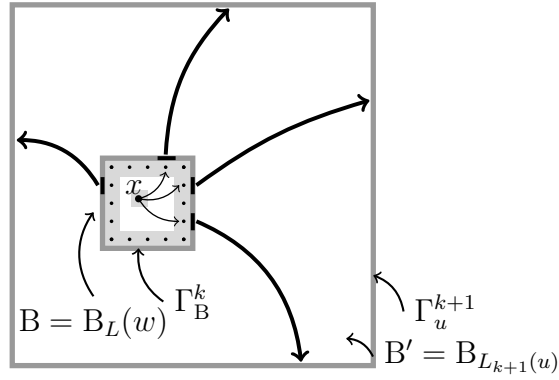
\begin{figure}
\begin{tabular}{c}
%
\begin{tikzpicture}
\begin{scope}[scale=0.12]
\clip (-44,-22) rectangle ++(90.0, 45.0);

\draw[color=black!40!white, line width = 2] (-20, -20) rectangle ++ (40, 40);

\foreach \i in {-10, -8, ..., -2}
{
  \fill[color=gray!30!white]  (\i, -7) rectangle ++(1.9, 1.9);
  \fill[color=gray!30!white]  (\i, 1) rectangle ++(1.9, 1.9);
}

\foreach \i in {-7, -5, ..., 1}
{
  \fill[color=gray!30!white]  (-10, \i) rectangle ++(1.9, 1.9);
  \fill[color=gray!30!white]  (-2, \i) rectangle ++(1.9, 1.9);
}

\foreach \i in {-10, -8, ..., -2}
{
  \fill (\i+1, -7+1) circle (0.2);
  \fill (\i+1, 1+1) circle (0.2);
}

\foreach \i in {-7, -5, ..., 1}
{
  \fill  (-10+1, \i+1) circle (0.2);
  \fill  (-2+1, \i+1) circle (0.2);
}

\fill[color=gray!30!white]  (-6.8, -2.4) rectangle ++(1.8, 1.8);

\draw[color=black!40!white, line width = 2] (-10, -7) rectangle ++ (10, 10);

\fill[color=black] (-6, -1.5) circle (0.4);


\draw[->,line width=0.5, bend right=30] (-6, -1.5) to (-1.5,0);
\draw[->,line width=0.5, bend right=30] (-6, -1.5) to (-3,1.5);
\draw[->,line width=0.5, bend right=30] (-6, -1.5) to (-1.5,-4.0);
\node (x) at (-6.5, -0.2) {$x$};

\node (ball) at (-11.0, -15.3) {$\ball=\ball_{L}(w)$};
\draw[->,line width = 0.8, bend left=35] (ball.north) to (-11.0, -3);

\node (rGamk) at (-0, -12.9) {$\rGam^{k}_\ball$};
\draw[->,line width = 0.8, bend left=30] (rGamk.west) to (-6, -7.0);

\node (ballp) at (31, -19) {$\ball'=\ball_{L_{k+1}(u)}$};
\draw[->,line width = 0.8, bend right=30] (ballp.west) to (17, -17);

\node (rGamkone) at (28, -14) {$\rGam^{k+1}_u$};
\draw[->,line width = 0.8, bend right=30] (rGamkone.west) to (20, -10);

\draw[line width = 1.5] (-3.9, 3) to (-1.9,3);
\draw[->,line width=1.5, bend left=20] (-3,3.4) to (4.0,19.57);

\draw[line width = 1.5,color=black] (0.0, -1.0) to (-0.0, 0.9);
\draw[->,line width=1.5, bend left=10] (0.4,-0.0) to (19.57,9.4);

\draw[line width = 1.5,color=black] (0.0, -5.0) to (-0.0, -3.1);
\draw[->,line width=1.5, bend left=30] (0.4,-4.0) to (12.0,-19.57);

\draw[line width = 1.5,color=black] (-10.0, -1.0) to (-10.0, +0.9);
\draw[->,line width=1.5, bend right=30] (-10.5,0.0) to (-19.5,5.0);

\end{scope}
\end{tikzpicture}
%
\end{tabular}

\caption{
\footnotesize
Example for the inequality \eqref{eq:bound.GRI.dist.Sigma.general}.
Black dots are the centers $c\in \cC^k(\rGamk_\ball)$ of admissible cells covering
the set $\rGamk_\ball$, shown as a dark gray annulus.
Long thick arrows represent the terms $\bGam^{k+1}_u G_{\ball'} \chik_c$
with $c\in \cC^k(\rGamk_\ball)$.
In \eqref{eq:bound.GRI.dist.Sigma.concentric}, $x=u$, so $\ball$ and $\ball'$ are concentric.
}
\end{figure}%

In Appendix \ref{app:domin.decay}, we will use \eqref{eq:bound.GRI.dist.Sigma.general}
in the situation where $x=u$, thus $\ball = \ball_L(u)$ is concentric with  $\ball'=\ball_{L_{k+1}}(u)$,
and $\dist\big(E, \Sigma( H_{\ball})\big) =: \eps_\ball>0$,  so
\be\label{eq:bound.GRI.dist.Sigma.concentric}
\bal
\big\| \bGam^{k+1}_{u} G_{\ball_{L_{k+1}}(u)} \chik_{c_x} \big\|
&\le \frac{\CWk }{\eps_\ball }
    \max_{c\in \cCk(\rGamk_\ball)} \big\| \bGam^{k+1}_{u} G_{\ball'} \chik_c \big\| ~.
\eal
\ee


\section{Dominated decay and EVC bounds. Main results}
\label{sec:dominated}

Below the centers of various cubes $\ball_{\bullet}(\bullet)$ will be assumed to
be the admissible centers at the respective scale.

Consider a cube $\ball = \ball_{L_{k+1}}(u)$ along with its skeleton graph $\cB$
defined in Section \ref{sec:basic}
(with the vertex set $\ball\cap\cC^k$),
and introduce the function $f:\cB\to\DR_+$ given by
$$
\bal
& f: \, x \mapsto \big\| \bGamkone_u G_{\ball} \chik_x \big\| ~.
\eal
$$
Then by GRI \eqref{eq:GRI.1},
\be\label{eq:GRI.h}
\bal
f(x) &\le \dnorm{ G_{\ball_{L_k}(x)}} \, \max_{ \lr{c, x}\in \cCk } \, f(c)  ~.
\eal
\ee

An inequality of the form \eqref{eq:GRI.h} is most useful when $\dnorm{ G_{\ball_{L_k}(x)}}  \le q < 1$;
in this case, using an iterated application of the GRI, it is not difficult to prove
the bound $f(u) \le  q^{ Y - 1}$.
Below we formulate a more advanced analog of this simple bound  (Lemma \ref{lem:domin.decay})
in a more general situation where for some $S\ge 1$,
there are at most $S$ vertices
$\rc\in\cB$ where $\dnorm{ \ball_{L_k}(\rc)}$ fails to be small.

Recall that $\Sigma_{u,L}$ stands for the spectrum of $H_{\ball_{L}(u)}$, and
$Y_{k}$ is an odd integer, so $Y_k = 2K_k + 1$.

\bde
\label{def:CNR}
Let be given an integer $k\ge 0$ and real numbers $\eps>0$ and $E$.

$\bullet$
A cube $\ball_{L}(u)$ is called $(E, \eps)$-NR (non-resonant), iff $\dist\left( \Sigma_{u,L}, E \right) \ge \eps$;

$\bullet$
A cube $\ball_{L_{k+1}}(u)$ is called $(E, \eps)$-CNR (completely non-resonant), iff for all
$j = K_k, \ldots, K_k + Y_{k+1} -2$ the cube $\ball_{j L_{k}/3}(u)$ is $(E,\eps)$-NR.
\ede

The role of the concentric cubes
$\ball_{j L_{k}/3}(u)$ will become clear in
Appendix \ref{app:domin.decay}, in the proof of Lemma \ref{lem:domin.decay}
(cf. also Fig.~4).

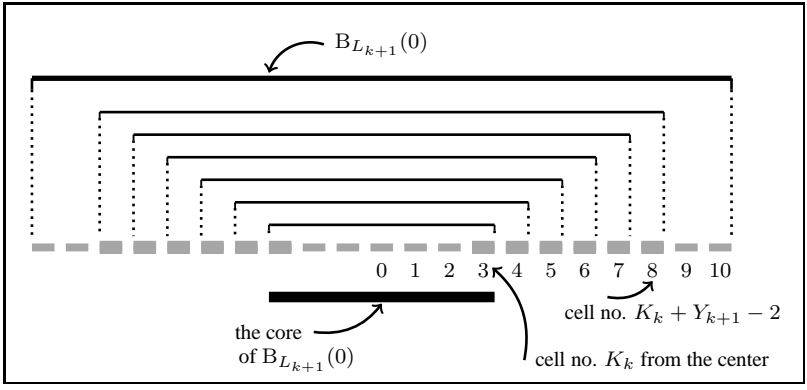
\begin{figure}
\begin{tabular}{|c|}
\hline
\begin{tikzpicture}

\begin{scope}[scale=0.30,yshift=400]
\clip (-16.0, -4.5) rectangle ++(34.0, 16.3);
\draw[color=white, line width = 1] (0, 12) -- (1, 12);

\foreach \i in {-10, -9, ..., 10 }
{
  \draw[color=black!70!white!50, line width = 3] (1.5*\i-0.5, 1.0) -- (1.5*\i+0.5, 1.0);
}

\draw[color=black, line width = 2] (- 1.5*10 - 0.5, 8.5) -- (1.5*10 +0.5, 8.5);

\draw[line width = 1] (-1.5*10-0.5, 8.5)  -- (-1.5*10 - 0.5, 8.0) ;

\draw[line width = 1] (1.5*10+0.5, 8.5)  -- (1.5*10 + 0.5, 8.0) ;
\draw[dotted, line width = 1] (-1.5*10-0.5, 8.0)  -- (-1.5*10 - 0.5, 1.5) ;
\draw[dotted, line width = 1] (1.5*10+0.5, 8.0)  -- (1.5*10 + 0.5, 1.5) ;

\node (ballkplusone) at (0, 10.0) {\scriptsize$\ball_{L_{k+1}}(0)$};
\draw[->, line width = 1, bend right=45] (ballkplusone.west) to (-5.0, 8.75);

\foreach \i in {-8, -7, ..., -3, 3, 4, ..., 8 }
{
  \draw[color=black!70!white!50, line width = 5] (1.5*\i-0.5, 1.0) -- (1.5*\i+0.5, 1.0);
}

\foreach \i in {0, 1, ..., 10 }
{
    \node at (1.5*\i, 0.0) {\scriptsize{$\i$}};
}

\node (jY) at (12, -4) {\scriptsize \text{cell no. $K_{k}$ from the center}};

\node (core) at (-5, -2.75) {\scriptsize \text{the core}};
\node (coreof) at (-3.75, -4) {\scriptsize \text{of $\ball_{L_{k+1}}(0)$}};

\draw[->, line width = 1, bend right =30] (core.east) to (0.0, -1.5);

\node (jYp) at (12.9, -2) {\scriptsize \text{cell no. $K_{k}+Y_{k+1}-2$}};

\draw[->, line width = 1, bend right =30] (jY.west) to (4.9, 0.1);
\draw[->, line width = 1, bend right =30] (10.0, -1.2) to (12.0, -0.5);

  \draw[color=black, line width = 4] (- 1.5*3 - 0.5, -1.2) -- ( 1.5*3 + 0.5, -1.2);
  \foreach \j in { 3, ..., 7, 8}
  {

  \draw[color=black, line width = 1] (- 1.5*\j - 0.5, \j - 1.0) -- (1.5*\j +0.5, \j-1.0);

  \draw[color=black, line width = 0.7] ( - 1.5*\j - 0.5, \j - 1.0) -- ( - 1.5*\j -0.5, \j -0.0-1.25);
  \draw[color=black, line width = 1, dotted] (- 1.5*\j - 0.5, \j - 1.25) -- ( - 1.5*\j -0.5, 1.5);
    \draw[color=black, line width = 1, dotted] ( + 1.5*\j + 0.5, \j - 1.25) -- (+ 1.5*\j +0.5, 1.5);

  \draw[color=black, line width = 0.7] ( + 1.5*\j + 0.5, \j - 1.0) -- ( + 1.5*\j +0.5, \j - 1.25);

  }
\end{scope}
\end{tikzpicture}
\\
\hline
\end{tabular}
\caption{
\footnotesize
A $1$-dimensional example for Definition \ref{def:CNR}. Here $K_k=3$, $Y_{k+1}=2 \cdot K_{k}+1 = 7$. The thick black interval
shows the position of the core of the cube $\ball_{L_{k+1}}(0)$, with some $k\ge 0$  and $L_{k+1} = Y_{k+1} L_k$,
decomposed into cells of size $\ell_k = L_k/3$ (short gray intervals). The CNR property requires the
$Y_{k+1}-1 = 6$ long black intervals (cubes, for $d>1$), to be non-resonant. These cubes may be used in the geometrical
procedure in the proof of Lemma \ref{lem:domin.decay} (cf. Appendix \ref{app:domin.decay}), along with
the cells shown as thick gray intervals. Thinner gray intervals are the remaining $\ell_k$-cells.
}
\end{figure}%

\bde
\label{def:NS}
Let be given an integer $k\ge 0$ and real numbers $\eps>0$ and $E$.
A cube $\ball_{L_k}(u)$ is called $(E, \eps)$-NS (non-singular), if $E \not\in \Sigma_{u,L}$, and
$\dnorm{ G_{\ball_{L_k}(u)}} \le \eps$. Otherwise, it is called $(E, \eps)$-S (singular).
\ede

Below we choose the sizes $L$ of cubes $\ball_L(u)$ and
the parameter $\eps>0$ figuring in Definitions \ref{def:CNR} and \ref{def:NS}
in a specific way. First, we take $L \in \{L_k, \, k\ge 0\}$, with $L_{k} = Y_{k} L_{k-1}$
and $Y_k$ defined in \eqref{eq:def.Y.k};
$\eps = L_k^{-b_k}$ in the context of Definition \ref{def:CNR},
while in the property $(E, \eps)$-CNR we set
$\eps = L_k^{-s_k}$, with recursively constructed
sequences $(b_k, k\ge 0)$ and $(s_k, k\ge 0)$ (cf. Table \eqref{eq:table2b}).

Now we are ready to formulate the main results of the paper.
\btm\label{thm:Main.ESL}
Consider the random Hamiltonian $H(\om)$ of the form \eqref{eq:def.H}
and assume that the marginal PDF $F_V$ of an IID random potential $V(\cdot\,;\om)$
is H\"{o}lder-continuous of some order $\fbH\in(0,1]$.
Further, assume that for some $b_0> d/\beta$ and $L_0\in\DN$ satisfying
\be\label{eq:cond.L0.main.1}
\bal
L_0 & \ge L_0(\eta,\tau) :=
\max \left[ 11^{1/\tau^2}, \; 9^{\frac{4(6d+\eta)}{\eta}},
  \; p_0^{-\frac{8}{3 \eta} }, \; p_0^{-\frac{8}{b_0} } \right]  ,
\eal
\ee
with $\tau=\frac{1}{16 d}$ and $\eta = \half(\beta b_0 - d)$ (cf. \eqref{eq:def.eta}),
the following condition is fulfilled:
$$
\pr{ \ball_{L_0}(0) \text{\rm \;is $(E, L_0^{-b_0})$-S}} < (3Y_1-4)^{-2d} \equiv 529^{-d}.
$$
Then there exist  positive sequences  $(\delta_k)_{k\ge 1}$, $(\kappa_k)_{k\ge 1}$ such that
$
\lim\limits_{k\to+\infty} \delta_k  = \lim\limits_{k\to+\infty} \kappa_k = 1
$
and for all $k\ge 0$,
$$
\sup_{u\in\DZ^d} \; \pr{ \ball_{L_k} \text{\rm \; is $\left(E, \eu^{-(L_k)^{\delta_k}}\right)$-S}  } \le \eu^{-(L_k)^{\kappa_k}} .
$$
\etm

See Section \ref{sec:ESL} for explicit bounds on $\delta_k$ and $\kappa_k$.

Let $\csB_1$ be the set of all bounded Borel functions $\phi:\DR\to\DC$ with
$\| \phi \|_\infty \le 1$. By well-known techniques (cf. Sect.~\ref{sec:FEMSA.to.DL}), one can infer from Theorem
\ref{thm:Main.ESL} the following result.

\btm\label{thm:Main.EFC}
Under the hypotheses of Theorem \ref{thm:Main.ESL}, there is
a function $f:\,\DR_+ \to \DR_+$ which admits exponential scaling limit, viz.
\be
\lim_{L\to +\infty} \frac{ \ln \ln f(L)}{ \ln L} = 1 \,,
\ee
and such that for all $x,y\in\DZ^d$ and for any sufficiently large domain $\Lam \subseteq\DZ^d$
containing $x$ and $y$, one has
\be
\esm{ \sup_{\phi\in\csB_1} \big| \langle \one_x \,|\, \phi\left(H_{\Lam}\right) \,|\, \one_y \big| }
\le f([x-y|) \,.
\ee
\etm

The principal analytic tool used in the proof of Theorem \ref{thm:Main.ESL}
is the following
\ble
\label{lem:domin.decay}
Let the integer sequences $(Y_k)$, $(S_k)$, $(L_k)$ and positive real sequences
$(b_k)$, $(s_k)$ be defined as in \eqref{eq:table2a}--\eqref{eq:table2b}.
Fix some $k\ge 0$ and suppose that a cube $\ball_{L_{k+1}}(u)$
\begin{enumerate}[\rm(i)]
  \item[$\bullet$] is $(E,L_k^{-s_k})${\rm-CNR}, and

  \item[$\bullet$] contains no collection of $(S_{k+1}+1)$ pairwise disjoint
{\rm$(E,L_k^{-b_k})$-S} cubes of size $L_k$ with admissible centers $c\in\cCk$.
\end{enumerate}
Then one has
\be\label{eq:lem.domin.decay}
\dnorm{  G_{\ball_{L_{k+1}}(u)}(E) } \le  L_k^{d/8} L_k^{ - b_k(Y_{k+1} - 5 S_{k+1} - 1)}  .
\ee
\ele

See the proof in Appendix \ref{app:domin.decay}; like its counterpart from \cite{GK01}, it is
in essence a variant (or rather an adaptation) of a well-known argument
going back to \cite{DK89}.

As usual in the MSA, we also need an eigenvalue concentration (EVC)
estimate to bound the norm of the resolvent near the spectrum.

\ble\label{lem:Wegner.Holder}
Assume that the marginal probability distribution of an IID random potential $V$
is H\"{o}lder-continuous of order $\beta\in(0,1]$. Then for any cube of size $L$
one has
\be\label{eq:Wegner.Holder}
\pr{ \ball_{L}(u) \text{\rm \; is not {\rm$(E,L^{-s})$-NR}} } \le \Const L^d\, L^{- \fbH s}.
\ee
\ele

In the case where $V$ admits a bounded probability density, hence $\fbH=1$, this is the
classical result by Wegner \cite{W81}; see also a short proof, e.g., in \cite{CL90}.
A simple adaptation to H\"{o}lder-continuous (and more generally, continuous) marginal distributions, sufficient for
our purposes, can be found in \cite{C13b}, where it is shown that an EVC bound for the potentials
with Lipschitz-continuous marginal PDF $F_V$ can be automatically transformed
into its counterpart for the PDF with an arbitrary continuity modulus. Optimal Wegner bounds
have been proved earlier for various types of operators; cf., e.g., \cite{CH,CHK,KN13b}.


\section{Adaptive feedback scaling}
\label{sec:ASFS}

\subsection{Technical assumptions and some useful inequalities}
\label{ssec:AFS.tech}

In the recursive construction of the sequences $(b_k)_{k\ge 0}$ and $(s_k)_{k\ge 0}$,
mentioned in the previous section, the crucial input parameter is $b_0$. Given the marginal
distribution $F_V$ of the random potential $V:\DZ^d\times \Om \to \DR$,
which we assume H\"{o}lder-continuous of order $\beta\in(0,1]$ until Section \ref{sec:lower.reg},
we always assume that
$ b_0 > d/\fbH $
and introduce the scaling parameters
\begin{align}
\label{eq:def.eta}
   \eta &:= \half ( \fbH b_0 -d ) >0,
\\
\label{eq:def.s0}
   s_0 & := \frac{d}{\fbH}  + \frac{\eta}{\fbH}  \equiv  b_0 - \frac{\eta}{\fbH} .
\end{align}
The initial length scale $L_0$ is always assumed to be large enough, to satisfy
the explicit condition \eqref{eq:cond.L0.main.1}.
Further, set
\be\label{eq:def.Y.S.L.1}
\bal
Y_1 & = 9, \;\; S_1 = 1, \;\; a_1 = (3Y_1 - 4)^d.
\eal
\ee
We make a \textbf{crucial assumption},
\be\label{eq:def.prob.p_0}
p_0 := \pr{ \text{ $\ball_{L_0}(0)$ is not $(E,L_0^{-b_0})$-NS} } < a_1^{-2d} = \frac{1}{529^d},
\ee
and introduce the parameters $\th_0\in(0, 1/3)$ and $\sigma_0>0$ by letting
\begin{align}
\label{eq:def.sigma.0.first}
1 - \frac{ \ln a_1 }{ \ln p_0^{-1}} = \frac{1 + 3 \th_0}{2}~,
\;\;
\sigma_0 &= \frac{ \ln p_0^{-1}}{ \ln L_0 } ~.
\end{align}
The scale-free probability threshold in the RHS of \eqref{eq:def.prob.p_0}
is slightly better than $841^{-d}$ given in \cite{GK01}. This marginal modification
is due to a geometrical strategy of the proof of Lemma \ref{lem:domin.decay} which deviates
from that of an analogous argument in \cite{GK01}. It is clear, however, that
the importance of the scale-free probability bounds from \cite{GK01} goes far beyond the explicit numerical
estimates for specific lattices.

Further, introduce an integer
\be\label{eq:def.fK}
\fK = \fK(p_0, Y_1) := \min\{k\ge 1: \; (1+\th_0)^k \ge 2d/\sigma_0\} ,
\ee
and define the integer sequences $(Y_k)_{k\ge 1}$, $(S_k)_{k\ge 1}$, and
$(L_k)_{k\ge 1}$ as follows:
\begin{align}
\label{eq:def.Y.k}
 Y_{k} &=
    \begin{cases} Y_1=9, & k \le \fK,
      \\ \left\lfloor L_{k-1}^{ 1/8 }\right\rfloor, & k>\fK,
    \end{cases}
\\
\label{eq:def.S.k}
 S_{k}  &:=
   \begin{cases} S_1=1, & k \le \fK,
        \\ \left\lfloor \frac{1}{9} Y_{k}  \right\rfloor, & k>\fK .
   \end{cases}
\end{align}
Next, let
\be
\label{eq:def.N.k+1}
L_{k+1} := Y_{k+1} L_{k}, \;\;
\;\;
N_{k+1} = Y_{k+1} - 5 S_{k+1} - 1,
\;\; k\ge 0,
\ee
and define for $k\ge 0$ (cf. \eqref{eq:def.s0})
\be\label{eq:def.b.k}
b_{k+1} = \frac{4}{5} N_{k} b_k, \;\;\;
s_{k+1} = \frac{1}{2} b_{k+1}.
\ee

\subsection{Main formulae and relations between technical parameters}
\label{ssec:tables}

\noindent
The probability $p_0$ is small enough, viz.
\be\label{eq:cond.p0.tables}
p_0 := \pr{ \text{ $\ball_{L_0}(0)$ is not $(E,L_0^{-b_0})$-NS} } < 23^{-2d} ,
\ee
and the initial length scale $L_0$ is large enough; specifically, it suffices that
\be\label{eq:bound.L0.again}
L_0 \ge L_0(\eta,\tau) :=
\max \left[ 11^{1/\tau^2}, \; 9^{\frac{4(6d+\eta)}{\eta}},
  \; p_0^{-\frac{8}{3 \eta} } , \; p_0^{-\frac{8}{b_0} }  \right] .
\ee
It is to be emphasized that the parameters $\tau$, $\eta$, $p_0$ figuring in the hypothesis
\eqref{eq:bound.L0.again}
are related only to $d$ and $b_0$, so there is no vicious circle in
\eqref{eq:bound.L0.again}.

\renewcommand{\arraystretch}{2.0}
\be\label{eq:table2a}\hbox{\begin{tabular}{|c|c|}
  \hline
  $1 = S_1 < Y_1 = 9 $      &    $N_{1} := Y_1 - 5S_1 - 1  = 3  $
\\
  \hline
\rule{0pt}{1ex}
  $ b_0 > \frac{d}{\fbH} $
\rule[-1ex]{0pt}{0pt}
    & $ \eta = \shalf( \fbH b_0 - d) = \fbH s_0 - d >0 $
\\
  \hline
\rule{0pt}{4ex}
  $ a_1 = (3 Y_1 - 4)^d = 23^d$
  \rule[-3ex]{0pt}{0pt}
  & $ \rro_1 = \half \eta $, \;$\tau = \frac{1}{16 d}$
\\
  \hline
\rule{0pt}{2ex}
  $ 1 - \frac{\ln a_1}{\ln p_0^{-1}} = \frac{1+ 3 \th_0 }{2}, \; \th_0 <\frac{1}{3} $
  \rule[-2ex]{0pt}{0pt}
  & $ \fK = \min\left\{k:\, (1+\th_0)^k \ge \frac{2d}{\sigma_0} \right\} $
\\
  \hline
\rule{0pt}{2ex}
  $ \sigma_0 := \ln p_0^{-1}/\ln L_0  $
  \rule[-2ex]{0pt}{0pt}
  & $\tau_0 = \min\left[ \frac{\ln Y_1}{\ln L_0}, \, \frac{3\th_0}{1+3\th_0}, \, \tau \right]$
\\
  \hline
\end{tabular}}
\ee
\renewcommand{\arraystretch}{2.0}
\be\label{eq:table2b}\hbox{\begin{tabular}{|c|c|}
  \hline
\rule{0pt}{5ex}
     $ Y_{k+1} = \begin{cases} Y_1=9, & k \le \fK \\ \left\lfloor L_k^{ \tau }\right\rfloor, & k>\fK
     \end{cases} $
  \rule[-4ex]{0pt}{0pt}
 &
      $ S_{k+1} = \begin{cases} S_1=1, & k \le \fK \\ \left\lfloor \frac{1}{9} Y_{k+1} \right\rfloor, & k>\fK
     \end{cases} $
\\
\hline
     $ N_{k} = Y_k - 5 S_k - 1 \ge 3  $
    &
     $ \BA_{k} = \left( \frac{4}{5}\right)^k N_1 \cdots N_k $
\\
\hline
\rule{0pt}{5ex}
     $  \begin{array}{ll} b_{k}  = \BA_{k}b_0 = \frac{6}{5} s_k, \; k\ge 1
     \\
     a_k = (3Y_k - 4)^d \end{array} $
    &
     $B_k =
\begin{cases}
1+ \th_0 \in \big(1, \frac{4}{3} \big), & k \le \fK+1
\\
\frac{2}{3} (S_k + 1), & k > \fK + 1
\end{cases}$
\rule[-4ex]{0pt}{0pt}
\\
  \hline
   $\sigma_k  = \BB_k \sigma_0 = B_k \cdots B_1\, \sigma_0 $
   &
   $D_k =
\begin{cases}
\frac{4}{3}, & k \le \fK +1
\\
 \frac{2}{3} (S_k+1), & k > \fK +1
\end{cases}$
\\
  \hline
  $  \rro_{k}
  = D_{k} \cdots  D_2  \rro_1, \; k\ge 2  $ & $ \delta_{k}
  \ge  \frac{\ln(S_{k} + 1) - \ln(3/2)}{ \ln Y_{k}} \nearrow 1 $
\\
  \hline
  $p_{k} \le   L_{k}^{ - \sigma_{k} }, \; \sigma_k \le \rro_k $
   &
  $p_{k} := \pr{ \ball_{L_k}(u) \text{\rm\, is $(E, L_k^{-b_k})$-S } }  $
\rule[-2ex]{0pt}{0pt}
\\
  \hline
 $ \fw_{k} \le L_k^{-\rro_k} $ &
 $ \fw_{k} := \pr{ \ball_{L_k}(u) \text{ is  not $(E, L_{k+1}^{-s_{k}})$-CNR } } $
\\
  \hline
\end{tabular}}
\ee

\subsection{Unbounded  growth of the geometric scaling parameters}
\label{ssec:growth.Y.S}

The following statement is an important ingredient of the proof of exponential scaling limit
in the scheme with varying scaling parameters $Y_k, S_k$ (cf. Sect.~\ref{sec:ESL}).

\ble\label{lem:AFS.Y.S.n.integer}
Let be given an integer $L_0 \ge L_0(\eta)$, with $L_0(\eta)$ given by \eqref{eq:cond.L0.main.1}.
Then the sequences $(L_k)_{k\ge 0}$, $(Y_k)_{k > \fK}$  and $(S_k)_{k > \fK}$,  given by
\eqref{eq:def.Y.k}--\eqref{eq:def.S.k}, are strictly monotone increasing, and for all $k \ge 1$ one has
\be\label{eq:AFS.S.ge.Y.9}
 \frac{1}{10} Y_{k} \le S_k \le \frac{1}{9} Y_{k}.
\ee
\ele

\proof
Let
$L_0 \ge 11^{1/\tau^2}$, $k \ge \fK$.
Then we have $\tau_k = \tau$ and
$$
\bal
Y_{k+1} &= \Lf L_{k}^{ \tau } \Rf
=  \Lf Y_{k}^{ \tau } L_{k-1}^\tau  \Rf
=  \Lf \left(\Lf L_{k-1}^\tau \Rf \right)^{ \tau } L_{k-1}^\tau  \Rf
\ge  \Lf \left(\Lf L_{0}^\tau \Rf \right)^{ \tau } L_{k-1}^\tau  \Rf
\\
& \ge  \Lf \left(\Lf 11^{\tau/\tau^2} \Rf \right)^{ \tau } L_{k-1}^\tau  \Rf
\ge \Lf \left(\Lf 10^{\tau/\tau^2} + 1 \Rf \right)^{ \tau } L_{k-1}^\tau  \Rf
\\
& \ge\Lf \left(10^{\tau/\tau^2} \right)^{ \tau } L_{k-1}^\tau  \Rf
\ge 10 \Lf  L_{k-1}^\tau  \Rf = 10 Y_k > Y_k.
\eal
$$
Furthermore,
$$
\bal
S_{k+1} &= \Lf \frac{1}{9} Y_{k+1} \Rf  \ge \Lf 10 \cdot \frac{1}{9}  Y_k \Rf
\ge 10 \Lf \frac{1}{9}  Y_k \Rf  > S_k.
\eal
$$
Therefore, the sequences $(Y_k)_{k > \fK}$ and $(S_k)_{k > \fK}$ are strictly increasing.

\vskip2mm
To prove the LHS inequality in \eqref{eq:AFS.S.ge.Y.9}, notice that
$Y_{\fK+1} \ge 10 Y_{\fK}\ge 10\cdot 9$,
and for any real $y\ge 90$ one has
$
\left\lfloor \frac{y}{9} \right\rfloor \ge \frac{y}{9} - 1 \ge \frac{y}{10},
$
hence for all $k\ge \fK+1$
\be\label{eq:AFS.S.ge.Y.9.again}
  \frac{1}{10 } Y_k \le S_k = \left\lfloor \frac{1}{ 9 } Y_{k} \right\rfloor
\le \frac{1}{ 9 } Y_{k} ,
\ee
as asserted.
\qedhere

\subsection{Scaling of the GFs}

Lemma \ref{lem:good.CNR.is.NS.AFS} stated below is the usual analytic component of the MSA,
providing the conditions under which localization bounds at a scale $L_k$ imply
similar (or better, in this case) bounds at the next scale $L_{k+1}$. Its proof is very short, but
the bulk of technical work is actually done in b Lemma  \ref{lem:domin.decay}
proved in Appendix \ref{app:domin.decay}.

\ble\label{lem:good.CNR.is.NS.AFS}
Define the integer sequences $(Y_j)$, $(S_j)$, $(L_j)$, $(N_j)$ and positive real sequences
$(b_j)$, $(s_j)$ as in \eqref{eq:table2a}--\eqref{eq:table2b}.
Assume that
a cube $\ball_{L_{k+1}}(u)$, $k\ge 0$, is $(E, L_{k+1}^{-s_k})${\rm-CNR} and contains no collection of
$S_{k+1}+1$ pairwise disjoint $(E, L_k^{-b_k})${\rm-S} cubes of radius $L_k$
with admissible centers.
Then the cube $\ball_{L_{k+1}}(u)$ is  $(E, L^{-b_{k+1}})${\rm-NS}.
\ele
\proof
By Lemma \ref{lem:domin.decay}, we have
$
\bal
\dnorm{ G_{\ball_{L_{k+1}}}(u)}
 &\le  L_k^{ - b_k N_{k+1} + \frac{d}{8}} ~.
\eal
$
Therefore, recalling that $b_k > d/\fbH \ge d \ge 1$, $N_{k+1} \ge 3$, we obtain
$$
\bal
- \frac{\ln \dnorm{ G_{\ball_{L_{k+1}}}(u)} }{ \ln L_k} &
\ge b_k  N_{k+1} - \frac{d}{8} >  b_k  N_{k+1}\left( 1 - \frac{1}{8 N_{k+1}} \right)
\ge b_k \, \frac{ 23 }{24} N_{k+1} ~.
\eal
$$
With $L_{k+1} = L_k Y_{k+1}$,  $\ln Y_{k+1} / \ln L_k \le 1/(16 d) \le 1/16$,
we have
$$
\bal
b_k \frac{23}{24} N_{k+1} \cdot \frac{ 1 }{ 1 + \frac{1}{16} }
  > \frac{4}{5}  N_{k+1}\,  b_k= b_{k+1} ~.
\eal
$$
\qedhere


%
\subsection{Scaling of the probabilities}

In the next statement, we establish an important technical ingredient of the proof of the key
Lemma \ref{lem:scaling.prob}: an upper bound on the probability of ``tolerated resonances''
following from the Wegner estimate. As usual in the MSA, such upper bounds essentially shape
those on probability of ``insufficient'' decay of the Green functions and, ultimately, of
the eigenfunction correlators. Speaking informally, one cannot get bounds better than those stemming
from a Wegner-type analysis, so we have to
make sure the latter is compatible with the exponential scaling limit.

\ble\label{lem:scaling.Wegner}
Define the sequences of positive integers $(L_k)_{k\ge 0}$, $(Y_k)_{k\ge 1}$, $(S_k)_{k\ge 1}$
and positive real sequances
$(\fw_k)$, $(q_k)$, $(\rro_k)$
as in \eqref{eq:table2a}--\eqref{eq:table2b}.
Assume that $L_0 \ge 9^{\frac{12d+4}{\eta}}$ (cf. \eqref{eq:bound.L0.again})
Then the following bound holds
\footnote{Using the factor "$2$" in the RHS of \eqref{eq:assertion.w.k+1.AFS}
might seem artificial, but it
becomes convenient in \eqref{eq:rec.p.n.AFS}.}:
\begin{align}
\label{eq:assertion.w.k+1.AFS}
\forall\, k\ge 1 \qquad
  \fw_{k} &\le  q_{k} :=  2 L_{k}^{ - \rro_{k} } .
\end{align}
\ele
\proof
The probability $\fw_{k+1}$ refers to the event (cf. Definition \ref{def:CNR}) that for at least one integer
$j\in [ K_k, K_k + Y_{k+1} - 2 ]$, the cube $\ball_{j L_{k}/3}(u)$ is $(E,\eps)$-R. The probability
of each of these $Y_{k+1}-1$ events is bounded by Wegner estimate
(cf. \eqref{eq:Wegner.Holder}), thus
\be\label{eq:ln.w.start.Weg}
\bal
\frac{- \ln \left( \half \fw_{k+1} \right)}{ \ln L_{k+1} }
&\ge \frac{- \ln \left( Y_{k+1} \, L_{k+1}^d \, L_{k+1}^{ - \fbH s_{k} } \right)}{ \ln L_{k+1} }
\\
& = \fbH s_{k} \left( 1  - \frac{d}{ \fbH s_{k} }
   - \frac{ \ln( Y_{k+1}) }{ \fbH s_{k} \, \ln L_{k+1} } \right) .
\eal
\ee
%
\vskip1mm
\noindent
By hypothesis,
$L_0 \ge 9^{\frac{12d+4\eta}{\eta}} > \eu^{4 d/\eta}$,
so for $k=0$, \eqref{eq:ln.w.start.Weg} becomes
\be\label{eq:ln.w.n=0.Weg}
\bal
\frac{- \ln \left( \half \fw_{1} \right)}{ \ln L_{1} }
& \ge \fbH s_{0}  - d - \frac{ \varkappa_d }{ \ln L_{1} }
\ge  \eta - \frac{ 2 d }{ \ln L_{0} }
 \ge \half \eta = \rro_1;
\eal
\ee
the last RHS equality  is the definition of $\rro_1$ in \eqref{eq:table2a}.
Hence $\half \fw_{1} \le L_{1}^{ - \rro_{1}} = \half q_{1}$.

Now let $k\ge 1$.
By \eqref{eq:table2b},
$\fbH s_k \ge \fbH s_1 = \frac{5}{6}\frac{4 N_1}{5} \fbH b_0 = 2 \fbH b_0> 2d \ge 2$,
hence
$$
\bal
\frac{- \ln \left( \half \fw_{k+1} \right)}{ \ln L_{k+1} }
&\ge \fbH s_{k} \left( 1  - \frac{d}{ \fbH s_{1} }
  - \frac{ \ln Y_{k+1} }{ \fbH s_{1} \, \ln L_{k+1} } \right)
= \frac{5}{6} \fbH b_k \tgamma_{k} ~,
\eal
$$
where
$$
\bal
\tgamma_{k} & \ge
1 - \frac{1}{2}  - \frac{ \ln Y_{k+1} }{ 2 \ln L_{k+1} }
%
&
\ge  \frac{1}{2}  -  \frac{ \ln L^{1/(16d)}_{k} }{ 2 \ln L_{k} }
%
   > \frac{1}{4} ~,
\eal
$$
so $\fw_{k+1} \le 2 L_{k+1}^{-\fbH b_{k}/8} = 2 L_{k+1}^{ -\rro_{k+1} }$,
where $(\rro_j)$ follows the recursion
\be\label{eq:recurr.trro.k}
\frac{\rro_{k+1}}{\rro_k} = \frac{ b_k}{  b_{k-1}}
= \frac{4 N_{k}}{5} \ge \frac{12}{5} (S_k +1) > \frac{2}{3} (S_k +1) = D_{k+1} ~.
\ee
Since $\fw_1 \le q_1 = L_1^{-\rro_1}$, this implies
$\fw_j \le q_j = L_j^{-\rro_j}$ for all $j\ge 1$,
owing to the definition of $\rro_j = D_{j} \cdots D_2 \rro_1$ (cf. Table \eqref{eq:table2b}).
$\,$
\qedhere
\vskip1mm
Recall that we defined in \eqref{eq:def.fK} an integer
$\fK = \min\{k\ge 1: \, (1+\th_0)^k \ge \frac{2d}{\sigma_0}\}$. Now we turn to the heart
of the adaptive feedback scaling analysis.
\ble\label{lem:scaling.prob}
Consider the sequences of positive integers $(L_k)_{k\ge 0}$, $(Y_k)_{k\ge 1}$, $(S_k)_{k\ge 1}$
defined as in \eqref{eq:def.Y.k}--\eqref{eq:def.S.k}. Let $\{p_k,\, k\ge 0\}$ be defined as in
LHS equation \eqref{eq:assertion.p.n+1.AFS}.
%
Assume that $L_0 \ge 9^{\frac{12d+4\eta}{\eta}}$
(cf. \eqref{eq:bound.L0.again}),
and one has (cf. \eqref{eq:cond.p0.tables})
\begin{align}
\label{eq:lem.adap.Y.S.cond.p.n}
p_{0}  & <  (3Y_1 - 4)^{-2d} \equiv 23^{-2d}.
\end{align}
Define recursively a sequence of positive numbers $(\sigma_k)_{k\ge 0}$ :
\begin{align}
\label{eq:def.sigma.0}
 \sigma_0 &= \tsigma_0 = \ln p_0^{-1}/\ln L_0,
\\
\label{eq:def.sigma.k}
 \sigma_{k} &= \BB_k \sigma_0, \;\; \BB_k = B_1 \cdots B_k, \;\;
\\
\label{eq:def.B.j}
B_j &=
\begin{cases}
1+ \th_0 \in \big(1, \frac{4}{3} \big), & j=1, \ldots \fK
\\
\frac{2}{3} (S_j + 1) \ge \frac{4}{3}, & j\ge \fK+1 .
\end{cases}
\end{align}
Then for all $k\ge 1$, the following bound holds:
\begin{align}
\label{eq:assertion.p.n+1.AFS}
  p_{k} = \pr{ \ball_{L_k}(u) \text{\rm\, is $(E, L_k^{-b_k})$-S } } &=: L_{k}^{ - \tsigma_{k} }
  \le   L_{k}^{ - \sigma_{k} } ~.
\end{align}
\ele
\proof
By Lemma \ref{lem:good.CNR.is.NS.AFS}, if $\ball_{L_{k+1}}(u)$ is $(E,L_{k+1}^{-b_{k+1}})$-S,
then either it contains at least $(S_{k+1}+1)$ pairwise disjoint $(E,L_k^{-b_k})$-S cubes of size $L_k$,
or it is not $(E,L_{k+1}^{-s_{k}})$-CNR. Since $S_{k+1}+1\ge 2$, it follows by
a simple combinatorial calculation that, with $a_{k+1} = (3Y_{k+1}-4)^d$ \;\; (cf. \eqref{eq:table2b}),
\be\label{eq:rec.p.n.AFS}
\bal
p_{k+1}  &  \le \half \big( a_{k+1} p_k\big)^{S_{k+1} + 1} + \half \fw_{k+1} .
\eal
\ee
By Lemma \ref{lem:scaling.Wegner}, we have
$\fw_{k+1} \le q_{k+1} = L_{k+1}^{-\rro_{k+1}}$, thus
\be\label{eq:rec.p.n.q.n.AFS}
p_{k+1} \le \half (a_{k+1} p_k)^{S_{k+1}+1 } + \half q_{k+1}.
\ee

Now we shall analyze separately several cases: $\textbf{(A1)}$, $\textbf{(A2)}$, $\textbf{(B1)}$, and $\textbf{(B2)}$.

\vskip1mm
\noindent
$\textbf{(A1)}$
Suppose that for some $\kcirc \in[0, \fK-1]$ and all $k\in[0, \kcirc]$ one has
$p_{k+1} > q_{k+1}$.
(\emph{The analysis of the case (\textbf{A2}) below shows that once $p_{k+1} \le q_{k+1}$,
the inequality $p_{j} \le q_{j}$ then follows by induction for all $j\in[k+1, \fK]$.})
Observe that
\be\label{eq:ineq.p.k+1.q.k+1}
\half (a_{k+1} p_k)^{S_{k+1}+1 } \ge p_{k+1} - \half q_{k+1} \ge \half p_{k+1},
\ee
so
$p_{k+1} \le (a_{k+1} p_k)^{S_{k+1}+1 }$.
By finite induction in $k\in\{0, \ldots, \kcirc\}$ (which we are carrying out now), we know that
$\tsigma_k = \ln p_k^{-1}/\ln L_k \ge \sigma_0$: this is true for $k=0$ due to
\eqref{eq:def.sigma.0}, and for larger $k$ this will stem from the inequalities
\eqref{eq:ind.sigma.1}--\eqref{eq:ind.sigma.4} we are going to prove. Therefore,
starting with $k=0$, we can write
\be
\label{eq:ind.sigma.theta.0}
\bal
1 - \frac{d \ln a_{k+1} }{ \tsigma_{k} \ln L_{k}}
1 - \frac{d \ln a_{k+1} }{ \tsigma_{k} \ln L_{k}}
& \ge 1 -  \frac{d \ln a_{1} }{ \sigma_0 \ln L_{0}}  = 1 -  \frac{d \ln a_{1} }{ \ln p_0^{-1}}
= \frac{ 1 + 3 \th_0 }{ 2 } ~,
\eal
\ee
and with $S_{k+1}+1 \ge 2$, \; $\ln L_{k+1} = \ln L_k + \ln Y_{k+1}$, \, $Y_{k+1} = Y_1$,
we  obtain
\begin{align}
\label{eq:ind.sigma.1}
\frac{ \ln p_{k+1}^{-1} }{ \ln L_{k+1} } =
\tsigma_{k+1} &\ge
\tsigma_{k} \cdot (S_{k+1} + 1)
\left(1 - \frac{d \ln a_{k+1} }{\tsigma_{k} \ln L_{k}} \right) \frac{ \ln L_{k}}{ \ln L_{k+1} }
\\
\label{eq:ind.sigma.2}
& \ge \sigma_{k} \cdot (1 + 3 \th_0)
   \cdot \left( 1 -  \frac{ \ln Y_{1}}{ \ln L_{k} + \ln Y_{1} } \right)
\\
\label{eq:ind.sigma.3}
&  \ge \sigma_{k} \cdot (1 + 3 \th_0) \cdot \left(1 - \tau_0  \right)  \text{ \; (see ($\bigstar$) below)}
\\
\label{eq:ind.sigma.4}
& \ge \sigma_{k} \cdot (1 + \th_0) =B_{k+1} \sigma_k = \sigma_{k+1} \;\; \text{ \; (cf. \eqref{eq:def.B.j})},
\end{align}
so as long as $p_{k+1} > q_{k+1}$, we obtain inductively that $p_{k+1} \le L_{k+1}^{-\sigma_{k+1}}$, with
$$
\sigma_{k+1} = B_{k+1} \sigma_k, \;\;
B_{k+1} = 1+\th_0 < \frac{4}{3} = D_{k+1} \,.
$$
($\bigstar$) To derive \eqref{eq:ind.sigma.4} from \eqref{eq:ind.sigma.3}
we used the condition $\tau_0\le \frac{3\th_0}{1+3\th_0}$ from  \eqref{eq:table2a}.
\par\noindent
Thus \eqref{eq:ind.sigma.theta.0}, \eqref{eq:ind.sigma.1}--\eqref{eq:ind.sigma.4} hold inductively for all
$0 \le k \le \kcirc$, as claimed.

For further use, we need to make the following observations.

$\blacklozenge$
By \eqref{eq:def.sigma.0.first}, $\sigma_0 := \frac{\ln(p_0^{-1})}{\ln L_0}$,
and by \eqref{eq:table2a},  $\rro_1 = \frac{\eta}{2}$, thus
$$
\sigma_1 = B_1 \sigma_0 = (1+\th_0)\sigma_0 < \frac{4}{3} \sigma_0
= \frac{4}{3}\frac{\ln(p_0^{-1})}{\ln L_0} \le \frac{\eta}{2} = \rro_1\,,
$$
where the last inequality is equivalent to the assumption $L_0 \ge p_0^{-\frac{8}{3 \eta} }$
(cf. \eqref{eq:bound.L0.again}).

$\blacklozenge$ We also have $\sigma_2 \le \rro_2$, since
$$
\sigma_2 = B_2 B_1 \sigma_0 < \frac{16}{9} \sigma_0, \;\;
\rro_2 = \frac{1}{8} A_1 b_0 = \frac{12}{8 \cdot 5} b_0  = \frac{3}{10} b_0 \,,
$$
so $\sigma_2 \le \rro_2$ stems from
$$
\sigma_0 \le  \frac{9}{16} \cdot \frac{3}{10} \, b_0 \,,
$$
which, in turn, follows from the assumption $L_0 \ge p_0^{-\frac{8}{b_0}}$,
equivalent to $\ln p_0^{-1}/\ln L_0 \equiv \sigma_0 \le \frac{1}{8} b_0$
(cf. \eqref{eq:table2a}).

$\blacklozenge$
More generally, the inequality
\be\label{eq:note.sigma.j.le.rro.j}
\sigma_j \le \,\rro_j
\ee
follows inductively from $\sigma_2 \le \,\rro_2$ by the recursion $\sigma_j = B_j \sigma_{j-1}$,
$\rro_j = D_{j} \rro_{j-1}$, as long as $B_j \le D_{j}$, and the latter holds by \eqref{eq:table2b}:
$$
\bal
B_j &= 1+\th_0 < \frac{4}{3} = D_j, \;\; j\le \fK ~,
\\
B_j &= \frac{2}{3}\big( S_j +1) = D_j, \;\; j > \fK ~.
\eal
$$

The conclusion of this step of analysis is that for all $k\in[0, \kcirc]$,
$$
p_k \le L_k^{-\sigma_k}, \;\; \sigma_0 < \sigma_1 \le \sigma_k \le \,\rro_k .
$$

\vskip1mm
\noindent
$\textbf{(A2)}$
Consider the case where, for some $k < \fK$,
$$
L_{k+1}^{-\tsigma_{k+1} } = p_{k+1} \le q_{k+1} = L_{k+1}^{-\rro_{k+1} } ~.
$$
First, let us show that in this case $\left( a_{k+2} p_{k+1} \right)^{2} \le q_{k+2}$,
so by recursion \eqref{eq:rec.p.n.q.n.AFS} one has $p_{j} \le q_{j}$
for all $j = k+1, \ldots, \fK$.
Indeed, with $Y_{k+1} = Y_1 = 3^2$, $S_{k+1} = 1$ and
$$
a_{1} =(3Y_1-4)^d\le 3^d Y_{1}^{d} = Y_{1}^{3d/2} \,, \;\; L_{k+2} = Y_{1} L_{k+1} \,,
$$
the required inequality would stem from
$
\left( Y_{1}^{3d/2} p_{k+1} \right)^{2}
\le \left( Y_{1} L_{k+1} \right)^{-\rro_{k+2}}
$.
We have assumed $p_{k+1} \le q_{k+1} = L_k^{-\rro_{k+1}}$, so it would suffice that
$$
\bal
 Y_{1}^{3d} L_{k+1}^{-2\rro_{k+1}}
\le Y_{1}^{-\rro_{k+2}} L_{k+1}^{-\rro_{k+2}} ~,
\eal
$$
i.e.,
$(2\rro_{k+1} - \rro_{k+2}) \ln L_{k+1} \ge (\rro_{k+1} + 3d) \ln Y_1$.
Since $\rro_{k+2} = \frac{7}{4} \rro_{k+1}$, hence
$2\rro_{k+1} - \rro_{k+2} = \frac{1}{4} \rro_{k+1}$,
it suffices to validate the inequality
$$
\frac{1}{4} \rro_{k+1} \ln L_{k+1} \ge (\rro_{k+1} + 3d) \ln Y_1 ~,
$$
equivalent to
\be\label{eq:ineq.L.rro}
\ln L_{k+1} \ge \frac{4(\rro_{k+1} + 3d)}{\rro_{k+1} } \ln Y_1
= 4 \left( 1 + \frac{3d}{\rro_{k+1}} \right) \ln Y_1 ~.
\ee
Since $L_j\ge L_0$, $\rro_j\,\ge\, \rro_1 = \eta/2$, the relations
\eqref{eq:ineq.L.rro} \,stem from the hypothesis
$L_0 \ge Y_1^{ \frac{4(6d+\eta)}{\eta}} = 9^{ \frac{4(6d+\eta)}{\eta}}$, by a straightforward calculation.
As was said, by \eqref{eq:rec.p.n.q.n.AFS} this implies
\be\label{eq:A1.p.j.le.q.j}
\forall\, j\in\{1, \ldots, \fK\} \qquad L_j^{-\tsigma_j} = p_j \le q_j = L_j^{-\rro_j}
 \le L_j^{-\sigma_j} ~,
\;\; \text{ as } \sigma_j \le \,\, \rro_j \,.
\ee

The logic of the analysis for $k \ge \fK$ is quite similar to that of the cases $(\textbf{A1})$--$(\textbf{A2})$,
but here one has to operate with growing sequences $(Y_k)$, $(S_k)$.

\vskip1mm
\noindent
$\textbf{(B1)}$ Suppose that for some $\kcirc \ge \fK$ and all
$k\in [\fK, \kcirc]$, one has $p_{k+1} > q_{k+1}$.
By the same argument as in \textbf{(A1)} (cf. \eqref{eq:ineq.p.k+1.q.k+1}), it follows that
$p_{k+1} \le \left( a_{k+1} p_k\right)^{S_{k+1}+1}$.
Unlike the case \textbf{(A1)}, now
$\BB_k = B_k \cdots B_1 \ge  2d/\sigma_0$, by definition of $\fK$, hence
$$
\sigma_k = \BB_k \sigma_0 \ge  2d \,.
$$
As a preparation for the bounds \eqref{eq:ind.sigma.k+1}--\eqref{eq:ind.sigma.k+1.4},
recall
$Y_{k+1} = \lf L_k^{\tau_{k+1}} \rf \le L_k^{\tau_{k+1}}$,
$\tau_{k+1} = 1/(16d)$, $3Y_{k+1} < Y_{k+1}^2$,
so it follows that
$$
\bal
\frac{ \ln \big( (3Y_{k+1})^d \big) }{ \sigma_k \ln L_k }
< \frac{2d \ln Y_{k+1}}{ \sigma_k \ln L_k} & \le
\frac{2d \tau_{k+1} \ln L_{k}}{ \sigma_0 \BB_k \ln L_k} \le \frac{1}{8} ~,
\eal
$$
since
$\sigma_0\BB_k  \ge \sigma_0\BB_\fK \ge \sigma_0 (1+\th_0)^\fK \ge 2d$ (cf. \eqref{eq:def.fK}).
Therefore,
\begin{align}
\label{eq:ind.sigma.k+1}
\frac{ \ln p_{k+1}^{-1} }{ \ln L_{k+1} } =
\tsigma_{k+1} &\ge
(S_{k+1} + 1) \frac{\ln\left( a_{k+1} p_k\right)  }{\ln L_{k}} \frac{ \ln L_{k}}{ \ln L_{k+1} }
\\
& \ge \sigma_{k} \cdot (S_{k+1} + 1)
\left(1 - \frac{2d \ln Y_{k+1} }{\sigma_{k} \ln L_{k}} \right) \frac{ \ln L_{k}}{ \ln L_{k+1} }
\\
\label{eq:ind.sigma.k+1.2}
& \ge \sigma_{k} \cdot (S_{k+1} + 1)  \cdot \left( 1 -  \frac{1}{ 8 } \right)
   \frac{ \ln L_{k} }{ \ln L_{k}(1 + \tau_{k+1}) }
\\
\label{eq:ind.sigma.k+1.3}
&  \ge \sigma_{k} \cdot (S_{k+1} + 1)  \cdot \frac{7}{8} \cdot \frac{16}{17}
\\
\label{eq:ind.sigma.k+1.4}
& > \sigma_{k} \cdot \frac{2}{3}(S_{k+1} + 1) \ge  B_{k+1} \sigma_k ~.
\end{align}
Further, $N_j = Y_j - 5 S_j -1 \ge 9 S_j - 6S_j = 3 S_j$, and $1 + \th_0 < 4/3$,
thus
$$
D_j = \frac{4 N_j}{5} \ge \frac{12}{5} S_j > \frac{14}{9} S_j
\ge \max\left[ \frac{2}{3} (S_j +1), \, 1+\th_0\right] \ge B_j .
$$
\bre\emph{ A subtle (albeit merely technical) point here is the definition of $B_j$ and $D_j$ at $j=\fK+1$,
which may have attracted the reader's attention in Table \eqref{eq:table2b} and in \eqref{eq:def.B.j}:
switching from the initial values, resp., $1+\th_0$ and $4/3$, to the large ones growing with $j$,
is delayed by one step and occurs
at $j=\fK+1$. This is due to a one-step delay in the multiplicative recursion
$\rro_{j+1} = D_{j+1} \rro_j = \frac{4 N_j}{5} \rro_j$ in Lemma \ref{lem:scaling.Wegner} (cf. \eqref{eq:recurr.trro.k}).
Here we rely on the value $N_j$, not $N_{j+1}$. As such, this fact is not directly related to the
relation $\sigma_{\fK+1} = B_{\fK+1} \sigma_\fK$, but it is convenient to keep the general
bound $\rro_j \ge \sigma_j$, and this is why we set, artificially,
$B_{\fK+1} = B_\fK = 1+\th_0$, so as to have $D_{\fK+1} = \frac{4}{3} > B_{\fK+1}$.}
\ere

Now observe that
$\rro_{k+1}\ge \sigma_{k+1}$ for all $k$, since we have
(cf. \eqref{eq:recurr.trro.k})
$$
\frac{\rro_{k+1}}{\sigma_{k+1}}
= \frac{D_k}{B_k}\frac{\rro_{k}}{\sigma_{k}}
\ge \frac{\rro_{k}}{\sigma_{k}}
\ge \cdots \ge \frac{\rro_{1}}{\sigma_{1}} \ge 1 \,.
$$

\vskip1mm
\noindent
$\textbf{(B2)}$ Suppose that for some  $k\ge \fK$,
one has
$p_{k+1} \le q_{k+1}$.
To show by induction that $p_{j} \le q_{j}$ for all $j\ge k+1$,
we need to check, starting with $j=k+1$, that
\be\label{eq:p.j.q.j}
\left( a_{j} q_j \right)^{S_{j} + 1} \le q_{j+1},
\ee
for \eqref{eq:p.j.q.j} would yield immediately
\be\label{eq:p.j+1.q.j+1}
p_{j+1} \le \half \left( a_{j} q_j \right)^{S_{j} + 1} + \half q_{j+1}
\le \half q_{j+1} + \half q_{j+1} = q_{j+1} = L_{j+1}^{-\rro_{j+1}} ~.
\ee
The relation
\eqref{eq:p.j.q.j} is equivalent to
$$
\left(3^{2d} Y_j^{2d} \right)^{S_j +1} L_j^{ -(S_j+1) \rro_j}
\le Y_{j+1}^{-\rro_{j+1}} L_j^{ - \rro_{j+1}} \,.
$$
Further, on account of $3Y_{j} \le Y_j^{3/2}$ (as $Y_{j}\ge Y_1 = 3^2$), it suffices that
$$
 Y_j^{3d(S_j +1)} L_j^{ -(S_j+1) \rro_j}
\le Y_{j+1}^{-\rro_{j+1}} L_j^{ - \rro_{j+1}} \,.
$$
Recall $Y_i = L_{i-1}^{1/(16d)}$ for $i\ge \fK$, and $\rro_{j+1} = D_{j+1} \rro_j$.
By direct inspection, each of the inequalities \eqref{eq:B1.1}--\eqref{eq:B1.2} follows from the next one
in \eqref{eq:B1.1}--\eqref{eq:B1.3}:
\begin{align}
\label{eq:B1.1}
L_j^{ (S_j+1) \rro_j- \rro_{j+1}} &\ge  Y_j^{3d(S_j +1) + \rro_{j+1}}
\\
\label{eq:B1.2}
L_j^{ \big( (S_j+1) - D_{j+1} \big) \rro_j }
&\ge  L_j^{\big( \frac{3d}{16 d} (S_j +1) + \frac{1}{16 d} D_{j+1} \big)\rro_{j}}
\\
\label{eq:B1.3}
(S_j+1) \left(1 - \frac{3}{16} \right) &\ge \left( 1 + \frac{1}{16} \right) D_{j+1}
\end{align}
Finally the validity of  \eqref{eq:B1.3} follows from $D_{j+1} = \frac{2}{3}(S_{j+1} +1)$.
This proves \eqref{eq:p.j.q.j}, hence
\eqref{eq:p.j+1.q.j+1}.

The assertion of the lemma is proved.
\qedhere

\vskip3mm

$\blacklozenge$ This marks the end of the "renormalization group" analysis of the Green functions. In the next Section
\ref{sec:ESL} we interpret the finial outcome of this analysis as asymptotically exponential
decay of the GFs, and then derive in Section \ref{sec:FEMSA.to.DL} similar decay properties of the
eigenfunction correlators.

\section{Exponential scaling limit (ESL). Proof of Theorem \ref{thm:Main.ESL}}
\label{sec:ESL}

We have proved that for all $k\ge 0$,
\be\label{eq:final.decay.GF}
\pr{ \ball_{L_k} \text{ is $(E,L_k^{-b_k})$-S}  } \le L_k^{- \sigma_k} .
\ee
The language of ``polynomial'' bounds -- with growing exponents -- has been so far
convenient but certainly looked rather artificial,
so our next goal is to show that \eqref{eq:final.decay.GF} can be interpreted as follows:
$$
\pr{ \ball_{L_k} \text{ is $\left(E, \eu^{-(L_k)^{\delta_k}}\right)$-S}  } \le \eu^{-(L_k)^{\kappa_k}} ,
$$
where $\delta_k, \kappa_k \nearrow 1$ as $k\to+\infty$. This is a matter of simple calculations.

Indeed, by induction,
$b_{k} = \BA_{k} b_0$.
Since $S_j \le Y_j/9$, we have
\be\label{eq:Aj.Nj.Sj}
 N_j \ge Y_j - 5 S_j - 1
\ge 3 S_j \ge \frac{1}{3} Y_j .
\ee
Therefore,
\be\label{eq:how.s.k.grow}
\bal
b_k  &> b_0 3^{-k} \prod_{j=1}^k Y_j  = \frac{b_0}{L_0} 3^{-k} L_k ,
\\
& = L_k^{1 - \frac{1}{\ln L_k} \left( \ln \frac{L_0}{b_0} + k \ln 3\right)}
> L_k^{1 - \frac{\ln L_0 + 2k }{\ln L_k} } =  L_k^{1 - \ord{1}}
\eal
\ee
since $Y_j \nearrow +\infty$, thus $k/\ln L_k\to 0$. Consequently,
$$
L_k^{-b_k} \le \eu^{ - \ln L_k \cdot L_k^{1 - o(1)}} = \eu^{ - c_k \, L_k^{1 - o(1)}},
\quad c_k \tto{k\to+\infty} + \infty.
$$
More precisely, $L_k \approx L_{k-1}^{(16d+1)/(16d)}$ for $k> \fK$, so for some
$ 1 < q \approx \frac{16 d +1}{16 d}$,
$$
\ln L_k \ge C + C'  q^{k - \fK} \ge C'' q^k .
$$
Thus
$$
\frac{ \ln \ln L_k^{-b_k}}{ \ln L_k} \ge 1 - \frac{C'''}{(1 + \eps)^k},\;\; \eps \approx \frac{1}{16 d}.
$$

Similarly, for the probabilities $p_k \le L_k^{-\sigma_k}$ we have
$$
\ln p_k^{-1} \ge \sigma_k \ln L_k \ge \sigma_0 B_1 \ldots B_k,
$$
where $B_j \ge C S_j \ge C' Y_j$, $C, C'>0$, for all $j\ge \fK$. By taking a sufficiently small constant
$C''>0$, one can extend this lower bound to $B_1, \ldots, B_\fK$:
$$
\ln p_k^{-1} \ge C'' c^k Y_1 \cdots Y_k \ge C''' c^k L_k \ge L_k^{1 - \alpha(k)},
$$
with $\alpha(k) \le h^k$, $h\in(0,1)$.

%

\section{ESL for the eigenfunctions and their correlators}
\label{sec:FEMSA.to.DL}

It is well-known by now that a sufficiently fast decay of the Green functions, proved with sufficiently
high probability at each energy $E$ in a given interval $I\subseteq \DR$, implies both spectral localization
(a.s. pure point spectrum in $I$ with rapidly decaying eigenfunctions) and strong dynamical localization,
with rapidly decaying averaged EF correlators. Such implications can be established with the help of
different methods. For example, in the bootstrap method presented in Ref.~\cite{GK01}, the fixed-energy
estimates in probability, proved at a given energy $E_0$, are extended to an interval $I_0 = [E_0 - \eps, E_0+\eps]$
with sufficiently small $\eps>0$, by means of the energy-interval (a.k.a. variable-energy) MSA induction; the core
procedure goes back to earlier works \cite{Dr87,Sp88,DK89}.

In an earlier work \cite{C14a} (cf. also the book \cite{CS13}) we proposed an alternative approach based on an argument
employed by Elgart et al. \cite{ETV10} in the general context of the FMM and encapsulated in a fairly general,
abstract spectral reduction (FEMSA $\Rightarrow$ VEMSA). Similar ideas, in essence going back
to the work by Martinelli and Scoppola \cite{MS85}, were used in other papers; cf., e.g., \cite{BK05}.

Introduce the following notation:
\be
\BF_{x,L}(E,\om) = \max_{y\in \pt^- \ball_L(x)} \big| G_{\ball_L(x)}(x,y;E,\om) \big| .
\ee

We formulate the spectral reduction in the following way (cf. \cite{C14a,CS13}).
(\emph{Notice that the quantities $\bL$ are unrelated to the sequence of scaling exponents $b_k$.})

\btm
\label{thm:ETV}
Let be given a bounded interval $I\subset \DR$, an integer $L\ge 0$, two disjoint cubes $\ball_L(x)$, $\ball_L(y)$, and
the positive numbers $\aL, \fb_L, \fc_L, \fq_L$ satisfying
\be\label{eq:cond.a.b.c}
\fb_L \le \min \big[\aL \fc_L^2, \fc_L \big]
\ee
and such that
$$
\forall\, E\in I \qquad \max_{z\in\{x,y\}} \pr{ \BF_{z,L} > \aL } \le \fq_L .
$$
Assume also that, for some function $f:(0,1]\to \DR_+$,
\be\label{eq:thm3.Wegner}
 \forall\, \eps\in (0,1] \qquad
 \pr{ \dist\left(\Sigma(H_{\ball_L(x)}), \, \Sigma(H_{\ball_L(y)}) \right) \le \eps } \le f(\eps) \,.
\ee
Then
\be\label{eq:thm.ETV.pr.max}
\pr{  \sup_{E\in I}\; \max_{z\in\{x,y\}} \BF_{z,L}(E) > \aL }
   \le \frac{ |I|\, \fq_L }{ \bL } + f( 2 \ccL ) \,.
\ee
Consequently, taking into account the results of Section \ref{sec:ASFS},
for some $\delta_k \nearrow 1$ as $k\to +\infty$, one has
\be\label{eq:thm3.VEMSA}
\pr{\exists\, E\in I:\,  \text{ $\ball_L(x)$ and $\ball_L(y)$ are {\rm $(E,L_k^{-b_k})$-S} } }
\le \eu^{ - L^{\delta_k } } .
\ee
\etm

The proof given below is based on the following

\ble
\label{lem:for.thm.ETV}
Let be given $L\ge 0$ and positive numbers $\aL, \bL, \ccL, \fq_L$ such that
\begin{align}
\label{eq:cond.a.b.c.lemma}
\bL &\le \min \big[\aL \ccL^2, \ccL \big],
\\
\label{eq:pr.le.QL.lemma}
\sup_{E\in I} \;  &\pr{ \BF_{z,L}(E) > \aL } \le \fq_L .
\end{align}
There exists an event $\cB_z$ such that $\pr{\cB_z}\le \bL^{-1} \fq_L$ and for any
$\om\not\in \cB_z$, the set $\csE_z(2a) := \{E: \BF_{z,L}(E) > 2 \aL \}$
is contained in a union of intervals
$$
\bigcup_{j=1}^{M_z} I_j, \;\; I_j := \{ E:\, |E-E_j| \le 2\ccL \} \,, \;\
M_z \le (3L)^d,
$$
centered at the eigenvalues $E_j\in \Sigma(H(\om))\cap I$.
\ele
\proof
Consider the random subsets of the interval $I$ parameterized by $a'>0$,
$$
\csE(a';\om) = \{ E:\, \BF_{z,L}(E) \ge a' \}
$$
and the events parameterized by $b'>0$,
$$
\cB(b') = \{\om\in\Om:\, \mes(\csE(a) > b'\}
= \left\{\om\in\Om:\, \int_I \one_{\BF_{z,L}(E) \ge \aL}\, dE \; > b' \right\} .
$$
Using the hypotheses \eqref{eq:cond.a.b.c.lemma}-\eqref{eq:pr.le.QL.lemma},
apply Chebyshev's inequality and the Fubini theorem:
$$
\bal
\pr{ \cB(\bL) } &\le \bL^{-1} \esm{ \mes(\csE(\aL))}
\\
& = \bL^{-1} \int_I \, dE\, \esm{ \one_{\BF_{z,L}(E) \ge \aL} } \le \bL^{-1} \pr{ \BF_{z,L}(E) \ge \aL} .
\eal
$$
Fix any $\om\not\in \cB(\bL)$, so $\mes(\csE(\aL;\om))\le \bL$.

Further, consider the random sets parameterized by $c'>0$,
$$
\cR(c') = \{ \lam\in\DR: \, \min_j |\lam_j(\om) - \lam| \ge c'\}.
$$
Note that for $\aL\in(0,\ccL)$, $\cA_{\bL} := \{E:\, \dist(E, \cR(2\ccL))<\bL\}\subset \cR(\ccL)$,
hence $\cA_{\bL}^\rc=\DR\setminus \cA_{\bL}$ is a union of intervals at distance at least $\ccL$ from
the spectrum.

Let us show by contraposition that, for any $\om\not\in\cB(\bL)$, one has
$$
\{ E:\, \BF_{z,L}(E;\om) \ge 2\aL\} \cap \cR(2\ccL) = \vempty.
$$
Assume otherwise, and pick any point $\lam^*$ from the non-empty intersection in the LHS.
Let $J := \{E':\, |E-\lam^*| < \bL \}\subset \cA_{\bL} \subset\cR(\cL)$.
By the first resolvent identity
$$
\bal
\|G_{\ball_L(z)}(E')\| &\ge
\| G_{\ball_L(z)}(\lam^*) \| - |E'-\lam^*| \,\| G_{\ball_L(z)}(E') \| \, \| G_{\ball_L(z)}(\lam^*) \|
\\
& \ge 2\aL - \bL \cdot (2\fc_L)^{-1} (\fc_L)^{-1} \ge \aL,
\eal
$$
where in the last line we used the assumption \eqref{eq:cond.a.b.c}. We also used the bounds
$\|G_{\ball_L(z)}(\lam^*)\|\le (2\fc_L)^{-1}$ and
$$
\| G_{\ball_L(z)}(E') \| \le \frac{1} { \dist(E', \Sigma) }
\le \frac{1} { \dist(\lam^*, \Sigma) - |E' - \lam^*| }
\le \frac{1} {2\ccL - \bL },
$$
where $\bL \le \ccL$. Consequently, the entire interval $(\lam^* - \bL, \lam^* + \bL)$ of length
$2\aL >\bL$ is a subset of $\csE(\aL;\om)$, which is impossible for any $\om\not\in\cB(\bL)$.
This contradiction completes the proof.
\qedhere

For $\aL, \fc_L \le 1$, which is a frequent situation where Lemma \ref{lem:for.thm.ETV}
is applied, one can give a simpler variant of the bound \eqref{eq:thm.ETV.pr.max}.

\bco
\label{cor:ETV}
Let be given a bounded interval $I\subset \DR$, an integer $L\ge 0$ and disjoint cubes $\ball_L(x)$, $\ball_L(y)$.
Assume that for some $\aL, \fq_L\in(0,1]$
\be\label{eq:pr.le.QL.cor}
\sup_{E\in I} \; \max_{z\in\{x,y\}} \pr{ \BF_{z,L}(E) > \aL } \le \fq_L .
\ee
Assume also that, for some function $f:(0,1]\to \DR_+$,
\be
 \forall\, \eps\in (0,1] \qquad
 \pr{ \dist\left(\Sigma(H_{\ball_L(x)}), \, \Sigma(H_{\ball_L(y)}) \right) \le \eps } \le f(\eps)
\ee
Then
$$
\pr{ \sup_{E\in I}\; \max_{z\in\{x,y\}} \BF_{z,L}(E) > \max\left[ \aL, \, \fq_L^{1/2} \right]  }
   \le |I|\, \fq_L^{1/4}  + f\big( 2 \fq_L^{1/4} \big) \,.
$$
%
\eco
\proof
Let $\fa'_L = \max\big[ \aL, \, \fq_L^{1/2} \big]$, $\fc_L = \fq_L^{1/4}$,
then $\bL := \aL \fc_L^2 = \min[\fa'_L \fc_L^2, \fc_L]$, since $\fa'_L, \fc_L \le 1$.
Thus $(\fa'_L, \bL, \fc_L)$ fulfill the condition \eqref{eq:cond.a.b.c}. Further, the function
$s \mapsto \pr{ \BF_{z,L}(E) > s}$ is monotone decreasing, and $\fa'_L \ge \aL$, so
we have
$\pr{\BF_{z,L} > \fa'_L } \le \fq_L$,
and the claim follows from \eqref{eq:thm.ETV.pr.max}.
\qedhere

\vskip2mm
\noindent
\emph{Proof of Theorem \ref{thm:ETV}}
Define the events $\cB_x, \cB_y$ related to the points $x,y$
in the same way as the event $\cB_z$ relative to $z$ in the proof
of Lemma \ref{lem:for.thm.ETV}, and let $\cB = \cB_x \cup \cB_y$. Let
$\om\not\in \cB$. Then for both values of $z\in \{x,y\}$, the set
$\csE_z(a)$ is contained in the union of the intervals
$J_{z,i} = [E^{(z)}_i-2\ccL, E^{(z)}_i+2\ccL]$. Therefore,
$$
\pr{ \om:\, \inf_{ E\in I} \max_{z\in\{x,y\}} \BF_{z,L}(E) > \aL }
\le \bigpr{\om:\, \dist(\Sigma_x, \Sigma_y) \le 4\ccL } ;
$$
the latter probability is bounded with the help of \eqref{eq:thm3.Wegner}.
$\,$
\qedhere

\vskip2mm

Now the derivation of strong dynamical localization from the VEMSA-type estimates \eqref{eq:thm3.VEMSA}
can be obtained in the same way as in \cite{GK01}, directly in the entire lattice $\DZ^d$. This requires an a priori
polynomial bound of Shnol--Simon type (cf., e.g., \cite{Shnol57,Sim82})
on the growth rate of spectrally a.e. generalized eigenfunction;
the latter becomes unnecessary in arbitrarily large finite balls (cf. \cite{C14a,CS13,CS15a}).

\btm[Cf. {\cite[Theorem 7]{C14a}} ]
\label{thm:DL.GK}
Assume that the following bound holds true for some $\eps>0$, $L\in\DN$ and a pair of disjoint cubes
$\ball_L(x), \ball_L(y)$:
$$
\bigpr{ \exists\, E\in I:\; \text{$\ball_L(x)$ and $\ball_L(y)$ are $(E,\eps)${\rm-S} } }
\le \fh_L.
$$
Then for any cube $\ball_{L'}(w) \supset \left( \ball_{L+1}(x) \cup \ball_{L+1}(x) \right)$
one has
\be\label{eq:DL.GK}
\esm{ \big| \langle \one_x \,|\, \phi\left(H_{\ball}\right) \,|\, \one_y \big| } \le 4 \eps + \fh_L.
\ee
\etm

We had incorporated in the definition of the non-singular cubes, through the "norm" $\dnorm{\cdot}$,
a combinatorial factor measuring the volume of a cube $\ball_L(\cdot)$; otherwise, such a volume
factor would be present in the term $4\eps$  in the RHS of \eqref{eq:DL.GK}.

The extension of the EFC decay bounds to the entire lattice can be done with the help of the
Fatou lemma on convergent measures; such a path was laid down
in earlier works by Aizenman et al. \cite{Ai94,ASFH01,AENSS06}.

It is readily seen that the assertion of Theorem \ref{thm:Main.EFC} follows from
Theorem \ref{thm:Main.ESL} with the help of Theorem \ref{thm:DL.GK}.

Summarizing, one can say that the essential equivalence of various forms of Anderson
localization (decay of the GFs, EFs, EFCs) is firmly established by now for a large class
of random Hamiltonians.

\section{Lower regularity of the disorder}
\label{sec:lower.reg}

\btm
The results of Section \ref{sec:ASFS} remain valid for the marginal probability distributions
with continuity modulus $\fs_V(\cdot)$ satisfying the following condition:
for some $C'\in(0,+\infty)$ and an appropriately chosen $C>0$, for all
$\eps\in(0,1/2)$
\be
\label{eq:weak.cond.sV}
\fs_V(\eps) \le C' \eps^{ \frac{C}{\ln |\ln \eps|}} .
\ee
\etm

\proof

Consider first the case where $Y_{k+1} = \Lf L_k^{\tau} \Rf$, hence
$L_{k+1} \ge C L_k^{1+\tau}$, $\tau>0$.

The regularity of the marginal distribution of the random potential $V$
must be sufficient for proving a Wegner-type estimate
$$
\pr{ \|G_{\ball_{L_k}}(E)\| > L_k^{s_k}} \le L_k^{ - \beta_k s_k},
$$
where $\beta_k s_k$, replacing $\beta s_k$ used in the previous section,
has to be compatible with our main estimates. Denoting $\eps_k = L_k^{-s_k}$, we thus should have
$$
\pr{ \| G_{\ball_{L_k}}(E) \| > \eps_k } \le \eps^{ \beta_k} .
$$
Up to some inessential factors (depending on $L_k$), the above estimate can be inferred in a standard way
from the continuity of the marginal PDF $F_V$ with the continuity modulus of the form
$\fs_V(\eps) \le C \eps^{C' \beta_k}$.

Next, observe that one has $\eps_k^{-1} = L_k^{ s_k } \le \eu^{ c_1 L_k}$: indeed, our estimates by $L_k^{-b_k}$ and
$L_k^{\pm s_k}$ are not truly exponential in $L_k$ (although that would be very welcome), so
we only have $\eps_k \sim \eu^{ \pm L_k^{1 - o(1)} \ln L_k} = \eu^{ \pm L_k^{1 - o(1)} }$.
Thus
$$
\ln \ln \ln \eps_k^{-1} =
\ln \ln \ln L_k^{s_k} \le \ln \ln( c_1 L_k ) \le \ln  \ln( c_2 L_0^{q^k} ) \le c_3 k.
$$
At the same time, with $\beta_k = \frac{\beta_0}{(1+\kappa)^k}$, we have
$\ln \beta_k^{-1} \ge c_4 k$,
hence one can proceed with the scaling algorithm even in the case where
$$
\ln \beta_k^{-1} \ge c_5 \ln \ln \ln \eps_k^{-1}
\Longrightarrow \beta_k \le  \frac{c_6}{\ln |\ln \eps_k|  }.
$$
We conclude that the Wegner-type estimates compatible with the adaptive scaling scheme
employed in Section \ref{sec:ASFS}
can be inferred from the following condition upon the continuity modulus $\fs_V$:
$$
\fs_V(\eps)  \le C' \eps^{ \frac{ C }{ \ln | \ln \eps |} },
$$
which is -- just marginally -- weaker than H\"{o}lder regularity of any positive order.
Pictorially, it can be qualified as H\"{o}lder continuity of "almost zero" order.

The proof in the general case can be reduced to the above analysis, since
the double-exponential growth $L_k \sim L_0^{q^k}$ takes over the exponential one,
$L_k = L_0 Y_1^k$, after a finite number of steps $\fK = \fK(p_0)$.
Observe that
all intermediate calculations and bounds can be re-written in terms of strict
inequalities (for this is the case with the principal hypothesis, $p_0 < a_1^{-2}$),
and these strict inequalities can be preserved by replacing
$\beta = \Const$ with $\beta_k = Const/(1+\kappa)^k$ during the $\fK$ steps, provided
$\kappa>0$ is small enough -- depending of course on $\fK$.
The auxiliary constants
clearly depend upon the proximity of $p_0$ to the Germinet-Klein threshold $841^{-d}$.
After $\fK$ steps,
one can start the scaling procedure with $L'_0 := L_\fK$. In fact, this would be very close
in spirit to the Germinet-Klein first bootstrapping step.
\qedhere

\appendices

\section{Proof of Lemma \ref{lem:domin.decay}}
\label{app:domin.decay}

\bre
The reader may find the proof of the bound \eqref{eq:lem.domin.decay} given below excessively technical and opaque.
Indeed, we aim here to minimize the size of the "double-buffer" nonsingular zone around singular cubes;
this "buffering" technique essentially goes back to \cite{DK89}. However, a simple geometrical consideration
shows that, at least for some generous buffering size, resulting in some large constant, say, $A=100$,
one can achieve a non-optimal analog of \eqref{eq:lem.domin.decay} with the RHS of the form
$ L_k^{ - b_k(Y_{k+1} - A S_{k+1})}$. This would still be sufficient for the existence
of some scale-free probabilistic threshold $p_0\in(0,+\infty)$ for the onset of localization. Below some efforts are
made to keep $p_0 \le 29^{-2d}$, as the latter appears in the Germinet--Klein BMSA scheme from \cite{GK01}.
\ere

%
Fix $k\ge 0$ and consider the cube $\ball=\ball_{L_{k+1}}(u)$, where $L_{k+1} = Y_{k+1} L_k$,
$Y_{k+1} = 2K_{k+1}+1$, $L_k = 3\ell_k$,
and its $\ell_k$-skeleton graph $\cB$. For $r\ge 0$, let $\cB_r = \cB_r(u)$ be the
balls $\{c \in \cCk: \rd_{\cCk}(u,c) \le r\}$, and $\cL_r := \{c\in\cB: \rd_\cB(u,c)=r\}$
its boundary (a spherical layer).
Then $\cB = \cB_{R}$ with $R=3K_{k+1}+1$.
Recall that the vertices of $\cB$ represent the $\ell_k$-cells (or their centers) in the original lattice $\DZ^d$.

We will reduce our analysis of the function $\cB \ni c \mapsto |G_\ball(c,y;E)|$
to that of a monotone function of one integer variable
\be\label{eq:def.F.r}
F:\,  r \mapsto \max_{c: \,\rd_{\cB}(u,c) \le r} |G_\ball(c,y;E)| ;
r \in I := [0,  R ] \,.
\ee
We shall see that of particular interest is the sub-interval
$$
 I := [\,K_{k+1}, K_{k+1} + Y_{k+1}-2] = [R - Y_{k+1}, R - 2] \,,
$$
with $R = 3K_k+1$.
The reason is that this interval of radii corresponds to the shell of the cube $\ball_{k+1}(u)$
through which the GFs propagate, from the core to the boundary, and it is the decay across
this spherical layer that we have to assess. This analysis is essentially reduced to the
points with $r\ge R - Y_{k+1}$: for smaller radii, we apply first the GRI and jump to the boundary
of the core (with maximal penalty upper-bounded thanks to the CNR condition), and then start the analysis
presented below (cf. item \textbf{(C)}).

Introduce another function,
\be\label{eq:def.f.r}
f:\,  r \mapsto \max_{c \in\cL_r} |G_\ball(c,y;E)|
 \equiv \max_{c: \,\rd_{\cB}(u,c) = r} |G_\ball(c,y;E)| ~,
\ee
so that $F(r) = \max_{r'\le r} f(r')$.

Call a vertex $c\in \cB$ non-singular if the associated ball $\ball_{L_k}(c)\subset \DZ^d$
is $(E,L_k^{-b_k})$-NS, and singular, otherwise. Respectively, call $r \in I$
non-singular if all vertices $c$ with $\dist_\cB(u,c)=r$ are non-singular, and singular, otherwise.
The notions of singularity/non-singularity do not apply to $r\in[R-1,R-2]$.


By definition of the non-singular balls
(cf. Definition \ref{def:NS} and \eqref{eq:def.dnorm})
we have the following inequalities:

\noindent
\textbf{(A)} for any non-singular $r\in I$,
$$
f(r) \le \max_{r' \in[r-1, r+1]} \CWk^{-1} L_k^{-b_k} \,  f(r');
$$

\noindent
\textbf{(B)} owing to the assumed CNR-property of $\ball_{L_{k+1}}(u)$, for any $r\le r' \le R-2$,
an application of the GRI (cf. \eqref{eq:bound.GRI.dist.Sigma.concentric})
to the ball $\Lam_{r'L_k}(u)\subset\ball_{L_{k+1}}(u)$ gives
\be\label{eq:prop.B}
f(r) \le \CWk L_{k+1}^{s_k} \, f(r') \,.
\ee
To be more precise,
an application of the GRI is required for $r \le r'-1$, while for $r=r'$
the inequality \eqref{eq:prop.B}  follows trivially from $Y_{k+1}^d L_{k+1}^{s_k} \ge 1$.

Combining \textbf{(A)} and \textbf{(B)}, we come to the following statement:
\par
\noindent
\textbf{(C)} Assume that for some $r' \in I$, all points
$\rho \in [r'+3, r'+5]$ are non-singular. Then for all $r \in[0,r'+5]$
\be\label{eq:prop.C}
F(r) \le C_{W,k}^{-1}\,  L_k^{-2b_k} L_{k+1}^{s_k} \,  F(r'+6).
\ee

Notice that for $r = r'+5$, \eqref{eq:prop.C} follows immediately from the assumed non-singularity
of the point $r'+5$, so it remains to be established only for $r \le r'+4$.

For the proof, we first apply \textbf{(B)}:
\be\label{eq:B.first}
F(r'+4) =
\max_{ \rho \le r'+4} \;\;  f(\rho) \le \CWk  L_{k+1}^{s_k}  \,  f(r'+4).
\ee
Next, apply \textbf{(A)} to $r'+4$ (which is non-singular by assumption):
\be\label{eq:A.first}
f(r'+4) \le C_{W,k}^{-1}\, L_{k}^{-b_k} \, \max_{ r'' \in [r'+3, r'+5]} f(r''),
\ee
thus
\be\label{eq:B.first.A}
F(r'+4)
\le   \CWk^{-1}\, L_{k+1}^{s_k}  \,   L_{k}^{-b_k} \, \max_{ r'' \in [r'+3, r'+5]} f(r'').
\ee
Apply \textbf{(A)} once again to the three points $r''\in[r'+3, r'+5]$
(all of which are non-singular by assumption):
\be\label{eq:A.last}
\bal
\max_{ r'' \in [r'+3, r'+5]} \; f(r'')
& \le \CWk^{-1}\, L_{k}^{-b_k} \,
   \max_{ r'' \in [r'+3, r'+5]} \; \max_{ r''' \in [r''-1, r''+1]} f(r''')
\\
& \le \CWk^{-1}\, L_{k}^{-b_k} \,\; \max_{ r''' \in [r'+2, r'+6]} f(r''')
\\
& \le \CWk^{-1}\, L_{k}^{-b_k} \,\; F(r'+6).
\eal
\ee
Collecting \eqref{eq:B.first.A} and \eqref{eq:A.last}, the assertion \textbf{(C)} follows,
since $F(r) \le F(r')$ for $r\le r'$.

\begin{figure}
\begin{tabular}{|c|}
\hline
\begin{tikzpicture}

\begin{scope}[scale=0.28,yshift=400]
\clip (-1.2, -7.0) rectangle ++(35.0, 14.0);
\draw[color=white, line width=1] (0, 12) -- (1, 12);

\foreach \i in {0, 1, ..., 22 }
{
  \draw[color=black!40!white!50, line width=2] (1.5*\i-0.5, 1.0) -- (1.5*\i+0.5, 1.0);
}
\foreach \i in {0, 1, 7, 8, 9, 10, 16, 20, 21, 22 }
{
  \node at (1.5*\i, 0.0) {\scriptsize{$\i$}};
}

\foreach \i in { 10, 16, 17 }
{
  \draw[color=black, line width=4] (1.5*\i-0.5, 1) -- (1.5*\i+0.5, 1);
}

\foreach \i in {12, 13, 14, 15 }
{
}

\foreach \i in {10}
{

  \draw[color=black, line width=2] (1.5*\i + 1.5 - 0.5, -1.7) -- (1.5*\i+1.5*5+0.5, -1.7);

  \draw[line width=1] (1.5*7 - 0.5, -6.2) to (1.5*20 + 0.7, -6.2);
  \draw[dotted, line width=1] (1.5*20 +0.7 , -5.5) -- (1.5*20 + 0.7, 0.7);
  \draw[dotted, line width=1] (1.5*7 - 0.5 , -5.5) -- (1.5*7 - 0.5, 0.7);

  \node (I) at (1.5*14, -5.4) {\scriptsize $I =[\,K-1, K + Y-2] = [7, 20]$};

  \draw[line width=1] (1.5*0 , 4.5) to (1.5*7 + 0.7, 4.5);
  \draw[dotted, line width=1] (-2 , 4.5) to (1.5*0 + 0.7, 4.5);
  \draw[dotted, line width=1] (1.5*7 + 0.7 , 4.5) -- (1.5*7 + 0.7, 0.7);

  \node (core) at (1.5*2, 5.3) {\scriptsize\text{ the core }};
  \node (cent) at (1.5*2 + 1, 3.0) {\scriptsize\text{ the center }};
  \draw[->, line width=1, bend right = 45] (cent.west) to (0.0, 1.5);

  \draw[color=black!50!white, line width=2] (1.5*\i + 1.5*2 - 0.5, 4.0) -- (1.5*\i +1.5*4 +0.5, 4.0);
  \draw[color=black!50!white, line width=2] (1.5*\i + 1.5*3 - 0.5, 5.5) -- (1.5*\i +1.5*5 +0.5, 5.5);
  \draw[color=black!50!white, line width=2] (1.5*\i + 1.5*4 - 0.5, 3.0) -- (1.5*\i +1.5*6 +0.5, 3.0);


  \node at (1.5*\i + 1.5*3,   -0.5) {$\cJ_1$};

  \node at (1.5*\i + 1.5,   -1.0) { \scriptsize{$\alpha_1$} };
  \node at (1.5*\i + 1.5*5,   -1.0) { \scriptsize{$\beta_1$} };
  \node at (1.5*\i + 1.5*6+0.7,   -2.6) { \scriptsize{$\beta_1+1$} };
  \node at (1.5*\i + 1.5*0+0.7,   -2.6) { \scriptsize{$\alpha_1 - 1$} };

 \draw[->, bend right=25, line width=1] (1.5*10, -3.2) to (1.5*16, -3.2);

  \node (rprime) at (1.5*\i, 2.5) {};
  \draw[->, bend left = 70] (rprime.north) to (\i*1.5 + 1.5*4, 6.1);
  \node (rpc) at (1.5*\i+1.5*4, 5.2) {};
  \draw[->, bend right = 30] (rpc.west) to (\i*1.5 + 1.5*3, 4.2);
  \draw[->, bend left = 30] (rpc.east) to (\i*1.5 + 1.5*5, 3.2);

  \node (rpll) at (1.5*\i+1.5*3, 3.7) {};
  \draw[->, bend right = 30] (rpll.west) to (\i*1.5 + 1.5*2, 1.5);
  \draw[->, bend left = 30] (rpll.east) to (\i*1.5 + 1.5*3.9, 1.5);

  \node (rprr) at (1.5*\i+1.5*5, 2.8) {};
  \draw[->, bend right = 30] (rprr.west) to (\i*1.5 + 1.5*4.2, 1.5);
  \draw[->, bend left = 30] (rprr.east) to (\i*1.5 + 1.5*6, 1.5);
}

\foreach \i in {7}
{
  \draw[color=black, line width=1, dotted] (1.5*\i + 0.75, 0.25) -- (1.5*\i + 0.75, 1.75);
}

\foreach \i in {7, 8, 9, 16, 17, 18, 19, 20 }
{
  \draw[color=black!50!white, line width=4] (1.5*\i-0.5, 1) -- (1.5*\i+0.5, 1);
}

\node at (1.5*7,   2.5) {\scriptsize{$r_0$} };
\node at (1.5*8,   2.5) {\scriptsize{$r_1$} };
\node at (1.5*9,   2.5) {\scriptsize{$r_2$} };
\node at (1.5*10,  2.5) {\scriptsize{$r_3$} };

\node at (1.5*16+0.3,   2.5) {\scriptsize{$r_4$} };
\node at (1.5*20+0.0,   2.5) {\scriptsize{$r_8$} };
\node at (1.5*21+0.0,   2.5) {\scriptsize{$r_9$} };


%
\end{scope}

\end{tikzpicture}
\\
\hline
\end{tabular}
\caption{
\footnotesize
In this example, $K=7$, $Y=2 \cdot K+1=15$, $S=1$, $N = Y - 5S-1 = 9$, $R=22$.
It is assumed that there is no pair of disjoint singular intervals of the form $[\rho-1, \rho+1]$.
$r_3$ is the smallest integer in $I$ which is singular; it is the radial projection
of the center $c$ of a singular ball in the skeleton graph. It is this minimality property
which implies that $r_2 = r_3-1$ must be non-singular, despite the fact that the intervals
$[r_2-1, r_2+1]$ and $[r_3-1, r_3+1]$ overlap. On the other hand, due to the overlap of $[r_3-1, r_3+1]$
with $[r_3, r_3+2]$, the point $r_3+1$ may (or might) be singular, without producing a disjoint singular
pair. Therefore, we still can use the property \textbf{(A)} starting off the point $r_2$
(and aiming at $r_2+1=r_3$), but leaving from $r_3$, we have to make a longer flight
with possible "destinations" (i.e., reference points) ranging in
$\big[ (r_3+4)-2, (r_3+4)+2 \big]$$=$$[r_3+2,r_3+6]$. The longest flight consumes the distance
$6$, instead of $1$ that we would have for a non-singular departure point; this results in a loss
of $5$ points. The thick gray intervals indicate the "good" points providing the factors
$q\le (3Y-1)^{-d} L^{-b}<1$
in the "radial descent" recursion: $F(r_{i-1}) \le q F(r_i)$.
The point $r_9$ is used as the last reference point, but we can only bound $F(r_9)$
by the global maximum of $F$, since the GRI cannot be applied at a center $c$ of the skeleton
graph $\cB$ with $\rd_\cB(u,c) \ge R-1$.
In this example, we have the guaranteed decay bound
$F(r_0) \le q^9 F(r_9)\le q^9 F(R)$, with $9 = Y - 5S-1$.
}
\end{figure}
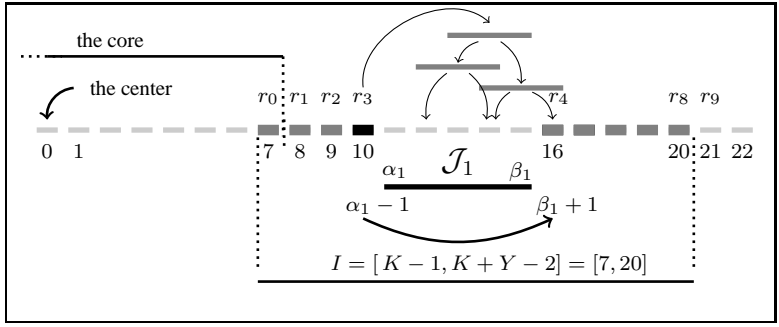%

Now pick any maximal collection of disjoint singular cubes $\ball_{L_k}(c_i)$,
$i=1, \ldots, n \le S$, denote $\rho_i = \rd_\cB(u,c_j)$,
and associate with each $c_i$
an interval $[\talpha_i,\tbeta_i] = [\rho_i -1, \rho_i + 5]$.
Next, decompose the union of intervals $[\talpha_i,\tbeta_i]$
into a disjoint union of maximal non-overlapping
intervals $\cJ_i=[\alpha_i, \beta_i]$, $1 \le i \le n' \le n$, so that
$\beta_i \le \alpha_{i+1}-1$; the equality $\alpha_{i+1} = \beta_i +1$ is permitted.

Note that for any $i$, all points $r\in[\beta_i-3, \beta_i]$ are non-singular,
otherwise we would have to augment $\cJ_i$ by fusing it with the interval $[r-1, r+5]$
overlapping with $\cJ_i$,
which contradicts the maximality of $\cJ_i$.

Let $I' = I \setminus \cup_i \cJ_i$ and enumerate the points of
$I'$ in the natural increasing order: $r_0, r_1, \ldots, r_M$. Next, add formally
the last point $r_{M+1}:=R-1$: it will be used as the last reference (destination) point, although
it cannot serve itself as a departure point.
If $r_i$ is non-singular, then we have $r_{i+1} = r_i +1$ and
$$
F(r_i) \le \CWk^{-1}\,\,L_k^{-b_k} F(r_{i+1})  ;
$$

otherwise, we can apply \textbf{(C)} and obtain
\be\label{eq:F.r.i}
\bal
F(r_i) &\le \CWk^{-1}\, L_k^{-2b_k} L_{k+1}^{s_k} F(r_{i+1})
= \|W\|^{-1} Y_k^{-d}\, L_k^{-2b_k} L_{k+1}^{s_k} F(r_{i+1})
\\
& \le \|W\|^{-1} \left( L_k^{-b_k + s_k} Y_{k+1}^{s_k}\right) \cdot Y_{k+1}^{- d}\, L_k^{-b_k} F(r_{i+1}),
\eal
\ee

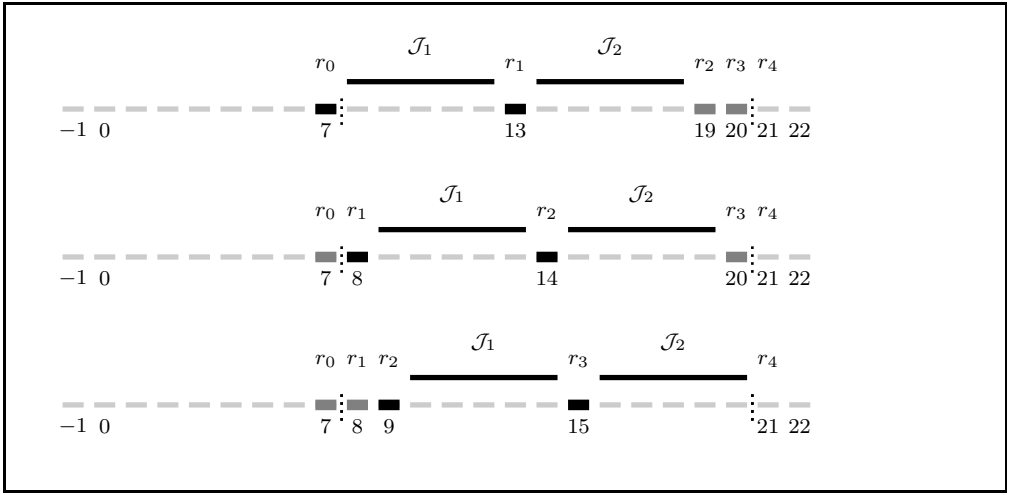
\begin{figure}
\begin{tabular}{|c|}
\hline
\begin{tikzpicture}


\begin{scope}[scale=0.28,yshift=400]
\clip (-3.0,-2.0) rectangle ++(45.0, 8.0);
\draw[color=white, line width=1] (0, 12) -- (1, 12);

\foreach \i in {-1, 0, ..., 22 }
{
  \draw[color=black!40!white!50, line width=2] (1.5*\i-0.5, 1.0) -- (1.5*\i+0.5, 1.0);
}
\foreach \i in {-1, 0, 7, 13, 19, 20, 21, 22 }
{
  \node at (1.5*\i, 0.0) {\scriptsize $\i$};
}

\foreach \i in {7, 13 }
{
  \draw[color=black, line width=4] (1.5*\i-0.5, 1) -- (1.5*\i+0.5, 1);
}

\foreach \i in {19, 20  }
{
  \draw[color=black!50!white, line width=4] (1.5*\i-0.5, 1) -- (1.5*\i+0.5, 1);
}

\foreach \i in {9 }
{
  \draw[color=black, line width=2] (1.5*\i-1.5*1-0.5, 2.3) -- (1.5*\i+1.5*3+0.5, 2.3);

  \node (Jaa) at (1.5*\i+ 1.5*1, 4) {\scriptsize$\cJ_1$};
}

\foreach \i in { 15 }
{
  \draw[color=black, line width=2] (1.5*\i-1.5*1-0.5, 2.3) -- (1.5*\i+1.5*3+0.5, 2.3);
  \node (Jab) at (1.5*\i+ 1.5*1, 4) {\scriptsize$\cJ_2$};
}

\foreach \i in {7, 20 }
{
  \draw[color=black, line width=1, dotted] (1.5*\i + 0.75, 0.25) -- (1.5*\i + 0.75, 1.75);
}

\node at (1.5*7,    3.1) {\scriptsize$r_0$};
\node at (1.5*13,   3.1) {\scriptsize$r_1$};

\node at (1.5*19,   3.1) {\scriptsize$r_2$};
\node at (1.5*20,   3.1) {\scriptsize$r_3$};

\node at (1.5*21,   3.1) {\scriptsize$r_4$};

\end{scope}

\begin{scope}[scale=0.28,yshift=200]
\clip (-3.0,-2.0) rectangle ++(45.0, 8.0);
\draw[color=white, line width=1] (0, 12) -- (1, 12);

\foreach \i in {-1, 0, ..., 22 }
{
  \draw[color=black!40!white!50, line width=2] (1.5*\i-0.5, 1.0) -- (1.5*\i+0.5, 1.0);
}

\foreach \i in {-1, 0, 7, 8, 14, 20, 21, 22 }
{
  \node at (1.5*\i, 0.0) {\scriptsize $\i$};
}

\foreach \i in {8, 14  }
{
  \draw[color=black, line width=4] (1.5*\i-0.5, 1) -- (1.5*\i+0.5, 1);
}

\foreach \i in {9 }
{
  \draw[color=black, line width=2] (1.5*\i-1.5*0-0.5, 2.3) -- (1.5*\i+1.5*4+0.5, 2.3);
  \node (Jba) at (1.5*\i+ 1.5*2, 4) {\scriptsize$\cJ_1$};
}
\foreach \i in { 15 }
{
  \draw[color=black, line width=2] (1.5*\i-1.5*0-0.5, 2.3) -- (1.5*\i+1.5*4+0.5, 2.3);
  \node (Jbb) at (1.5*\i+ 1.5*2, 4) {\scriptsize$\cJ_2$};
}

\foreach \i in {7, 20 }
{
  \draw[color=black, line width=1, dotted] (1.5*\i + 0.75, 0.25) -- (1.5*\i + 0.75, 1.75);
}

\foreach \i in {7, 20 }
{
    \draw[color=black!50!white, line width=4] (1.5*\i-0.5, 1) -- (1.5*\i+0.5, 1);
}

\node at (1.5*7,   3.1) {\scriptsize$r_0$};
\node at (1.5*8,   3.1) {\scriptsize$r_1$};
\node at (1.5*14,  3.1) {\scriptsize$r_2$};
\node at (1.5*20,  3.1) {\scriptsize$r_3$};
\node at (1.5*21,  3.1) {\scriptsize$r_4$};

\end{scope}


\begin{scope}[scale=0.28, yshift=0]
\clip (-4.0,-2.0) rectangle ++(45.0, 8.0);
\draw[color=white, line width=1] (0, 12) -- (1, 12);

\foreach \i in {-1, 0, ..., 22 }
{
  \draw[color=black!40!white!50, line width=2] (1.5*\i-0.5, 1.0) -- (1.5*\i+0.5, 1.0);
}
\foreach \i in {-1, 0, 7, 8, 9, 15, 21, 22 }
{
  \node at (1.5*\i, 0.0) {\scriptsize $\i$};
}

\foreach \i in {9, 15 }
{
  \draw[color=black, line width=4] (1.5*\i-0.5, 1) -- (1.5*\i+0.5, 1);
}

\foreach \i in {9 }
{
  \draw[color=black, line width=2] (1.5*\i+1.5*1 -0.5, 2.3) -- (1.5*\i+1.5*5+0.5, 2.3);
  \node (Jca) at (1.5*\i+1.5*3, 4) {\scriptsize$\cJ_1$};
}

\foreach \i in {15 }
{
  \draw[color=black, line width=2] (1.5*\i+1.5*1 -0.5, 2.3) -- (1.5*\i+1.5*5+0.5, 2.3);
  \node (Jcb) at (1.5*\i+ 1.5*3, 4) {\scriptsize$\cJ_2$};
}

\foreach \i in {7, 20 }
{
  \draw[color=black, line width=1, dotted] (1.5*\i + 0.75, 0.25) -- (1.5*\i + 0.75, 1.75);
  \draw[color=black, line width=1, dotted] (1.5*\i + 0.75, 0.25) -- (1.5*\i + 0.75, 1.75);
}

\foreach \i in {7, 8 }
{
  \draw[color=black!50!white, line width=4] (1.5*\i-0.5, 1) -- (1.5*\i+0.5, 1);
}

\node at (1.5*7,   3.1) {\scriptsize$r_0$};
\node at (1.5*8,   3.1) {\scriptsize$r_1$};
\node at (1.5*9,   3.1) {\scriptsize$r_2$};
\node at (1.5*15,  3.1) {\scriptsize$r_3$};
\node at (1.5*21,  3.1) {\scriptsize$r_4$};

\foreach \i in { 3}
{
}
\end{scope}

\end{tikzpicture}
\\
\hline
\end{tabular}
\caption{
\footnotesize
Here $K=7$, $Y=2 \cdot K+1=15$, $S=2$, $N = Y - 5S-1 = 4$. Three variants of the worst
case scenario, with non-overlapping intervals $\cJ_1$, $\cJ_2$.
The "useful" points of $I$, serving as departure points and providing a small factor are drawn as thick intervals
(black, if $r_i$ is singular, and dark gray, otherwise) and numbered as $r_0, \ldots, r_3$.
Here we have the guaranteed decay bound
$F(r_0) \le q^4 F(r_4) \le q^4 F(R)$. One can see that the number of useful points
becomes larger if two or more intervals $\cJ_i$ overlap with each other, or if at least one of them overlaps
with the complement of  $I = [R-Y, R - 2] $ $\,\big(\equiv [7,20]$, in this example$\big)$.
}
\end{figure}%

For $k=0$, we have $b_0 - s_0  = s_0 - \frac{d}{\beta} = \frac{\eta}{\beta}$, $\beta\in(0,1]$,
while $Y_1 = 9 < L_0$, so
$$
L_0^{-(b_0 - s_0)} Y_{1}^{s_0} = L_0^{-\frac{\eta}{\beta}} Y_{1}^{ \frac{d+\eta}{\beta} }
\le 1,
$$
owing to the assumption $L_0 \ge Y_1^{1 + \frac{d}{\eta}}$ (cf. \eqref{eq:cond.L0.main.1}).
For $k\ge 1$, we have $s_k = \frac{5}{6} b_k$, and again,
$$
L_k^{-b_k +s_k} Y_{k+1}^{s_k} = L_k^{-\frac{1}{6}s_k} Y_{k+1}^{s_k}
\le L_k^{-\frac{1}{6}s_k + \frac{1}{8} s_k}  < 1,
$$
hence for any $k\ge 0$,
\be\label{eq:F.r.i.again}
\bal
F(r_i) &\le \|W\|^{-1} L_k^{-b_k} F(r_{i+1}) ~.
\eal
\ee
Collapsing the intervals $\cJ_i$ removes from
$I$ at most $5 S_{k+1}$ points (see Fig. 6).
This upper bound becomes sharp if the radial projections of all
singular $L_k$-balls in the collection (fixed at the beginning) are non-overlapping.
Hence $|I'| \ge (Y_{k+1}-1) - 5 S_{k+1}=N_{k+1}$, so we obtain
\be\label{eq:bound.F.0}
\bal
 F(0) &\le C_{W,k}^{-1} L_k^{- b_k N_{k+1}} \le \frac{1}{\|W\|\, Y_k^d} L_k^{- b_k N_{k+1}}
 \le \frac{Y_{k+1}^d }{C_{W,k+1} }\, L_k^{- b_k N_{k+1}}
\\
&
\le L_{k}^{\tau_k d}\, C_{W,k+1}^{-1} \, L_k^{- b_k N_{k+1}}
\le L_{k}^{d/8}\, C_{W,k+1}^{-1} \, L_k^{- b_k N_{k+1}} ~.
\eal
\ee
We conclude that
\be
\bal
\dnorm{ G_{\ball_{L_{k+1}}}(u)} \le
C_{W,k+1} \, F(0) & \le L_{k}^{d/8}\,  \, L_k^{- b_k N_{k+1}} ~.
\eal
\ee
Lemma \ref{lem:domin.decay} is proved.
\qed

\section*{Acknowledgements}
I thank the Isaac Newton Institute, Cambridge, UK, and the organisers of the program
"\emph{Periodic and ergodic spectral problems}" (2015) for the support and opportunity to
work for six months in a stimulating atmosphere of the Institute, where the present work was completed.
It is also a pleasure to thank Alexander Elgart, Abel Klein, G\"{u}nter Stolz  and Yuri Suhov for fruitful
discussions.



\end{document}